\documentclass[conference]{IEEEtran}
\IEEEoverridecommandlockouts

\usepackage{amsmath,amssymb,amsfonts}
\usepackage{algorithmic}
\usepackage{graphicx}
\usepackage{textcomp}
\usepackage{xcolor}
\usepackage{lipsum}

\usepackage{dblfloatfix}

\usepackage{booktabs}
\usepackage{subeqnarray}
\usepackage{enumitem}
\usepackage{url}
\usepackage{subfig}
\usepackage{caption}
\usepackage[export]{adjustbox}
\usepackage{float}
\usepackage[square,numbers,sort&compress]{natbib}

\usepackage{eufrak}
\usepackage{siunitx}

\usepackage{hyperref}

\newcommand{\fvort}{\overline{\omega}}
\newcommand{\fpsi}{\overline{\psi}}
\newcommand{\fu}{\overline{\mathbf{u}}}

\usepackage{setspace}
\AtBeginDocument{%
  \addtolength\abovedisplayskip{-0.5\baselineskip}%
  \addtolength\belowdisplayskip{-0.5\baselineskip}%
  \addtolength\abovedisplayshortskip{-0.5\baselineskip}%
  \addtolength\belowdisplayshortskip{-0.5\baselineskip}%
}

\makeatletter
\newcommand{\linebreakand}{%
  \end{@IEEEauthorhalign}
  \hfill\mbox{}\par
  \mbox{}\hfill\begin{@IEEEauthorhalign}
}
\makeatother
\def\BibTeX{{\rm B\kern-.05em{\sc i\kern-.025em b}\kern-.08em
    T\kern-.1667em\lower.7ex\hbox{E}\kern-.125emX}}

\begin{document}

\title{\Large{Evaluation of Analytical Turbulence Closures for \\ Quasi-Geostrophic Ocean Flows with Coastal Boundaries}
\vspace{-11pt}}

\author{\IEEEauthorblockN{Anantha Narayanan Suresh Babu$^{*}$, 
Akhil Sadam$^{*}$, and
P. F. J. Lermusiaux$^{\dagger}$}
\IEEEauthorblockA{\textit{\textsuperscript{$ $}
Department of Mechanical Engineering, Massachusetts Institute of Technology, Cambridge, MA}\\
\IEEEauthorblockA{\textit{\textsuperscript{$ $}
Center for Computational Science and Engineering, Massachusetts Institute of Technology, Cambridge, MA}}}
\IEEEauthorblockA{\textsuperscript{$*$}Equal contribution}
\IEEEauthorblockA{\textsuperscript{$\dagger$}Corresponding author: pierrel@mit.edu}
\vspace{-11pt}
\vspace{-11pt}
} 

\maketitle
\thispagestyle{plain}
\pagestyle{plain}

\begin{abstract}
Numerical turbulence simulations typically involve parameterizations such as Large Eddy Simulations (LES). Applications to geophysical flows, especially ocean flows, are further complicated by the presence of complex topography and interior landforms such as coastlines, islands, and capes. In this work, we extend pseudo-spectral quasi-geostrophic (QG) numerical schemes and GPU-based solvers to simulate flows with coastal boundaries using the Brinkman volume penalization approach. We incorporate sponging and a splitting scheme to handle inflow and aperiodic boundary conditions. We evaluate four analytical sub-grid-scale (SGS) closures based on the eddy viscosity hypothesis: the standard Smagorinsky and Leith closures, and their dynamic variants. We show applications to QG flows past circular islands and capes with the $\beta$-plane approximation. We perform both a priori analysis of the SGS closure terms as well as a posteriori assessment of the SGS terms and simulated vorticity fields. Our results showcase differences between the various closures, especially their approach to phase and feature reconstruction errors in the presence of coastal boundaries.  
\end{abstract}


\section{Introduction}
High-resolution fluid simulations that fully resolve turbulence across all spatiotemporal scales are intractable in large geophysical domains such as the ocean and atmosphere \cite{pope2001turbulent}.
These flows range from basin-scales (thousands of kilometers) to the Kolmogorov scales (a few millimeters) \cite{vreugdenhil2019transport, kolmogorov1962refinement}. Since numerical simulations are limited in resolution, variable count, and modellable processes, they require additional parameterizations to account for unresolved scales and processes \cite{pope2001turbulent, gupta_lermusiaux_PRSA2021}. A common approach used for turbulence simulations is Large-Eddy Simulation (LES), which filters out subgrid-scale features, making some simulations tractable, though sometimes unreliable with solution divergence \cite{zhiyin2015large}. To ensure stable, reliable solutions, these simulations require accurate subgrid-scale (SGS) parameterizations \cite{armenio2002investigation,puthan2022wake,lermusiaux_2.29_notes}. 

Since LES performance is determined by the SGS parameterization, accurate SGS closure models have been actively researched since the 1960s \cite{jakhar2024learning}. SGS closure modeling began with the functional approach, based on the eddy viscosity hypothesis \cite{kraichnan1976eddy}. This approach introduces an artificial viscosity to capture dissipation from filtered fine-scales \cite{smagorinsky1963general, leith1968diffusion, baldwin1978thin}, and can be dynamic, where the eddy viscosity parameters are learned \emph{in real-time} \cite{germano1991dynamic, lilly1992proposed}. Another approach is the structural approach \cite{prakash2022invariant}, which directly reconstructs the SGS forcing via scale-similarity \cite{bardina1980improved} or local Taylor series expansions \cite{clark1979evaluation}. Mixed models combining the two approaches have also been explored \cite{leonard1999tensor}. Recently, data-driven and machine learning approaches have been developed \cite{zanna2020data, duraisamy2021perspectives}, including various deep learning SGS closure architectures using Artificial Neural Networks (ANNs) \cite{maulik2019subgrid}, Convolutional Neural Networks (CNNs) \cite{frezat2022posteriori, guan2022stable, srinivasan2024turbulence}, Neural Ordinary and Delay Differential Equations (NODEs and NDDEs) \cite{gupta_lermusiaux_SR2023}, and generative models like Generative Adversarial Networks \cite{perezhogin2023generative}.

In the ocean, resolving 100 m-10 km mesoscale and submesoscale dynamics \cite{mcwilliams2016submesoscale} is important to study energy, mass, and tracer transports. High-resolution satellite observations have also shown the presence of small coastal eddies and jets that dominate transport near capes \cite{digiacomo2001satellite}. However, direct LES application to the coastal ocean is hindered by the presence of complex topography, coastlines and interior landforms \cite{haidvogel2000coastal, roman2010large}. Previous works have evaluated analytical (MOLES, Mesoscale Ocean Large Eddy Simulations) \cite{graham2013framework} and \cite{ross2023benchmarking} benchmarked machine-learning closures in idealized open ocean flows. Data-driven closures have also been applied to global ocean models with coastlines, where these Convolutional Neural Networks (CNNs) based closures led to bias and unphysical artifacts near coastal boundaries \cite{guillaumin2021stochastic, zhang2023implementation}. Some artifact mitigation was explored in \cite{zhang2025addressing}.

In this work, we first extend pseudo-spectral quasi-geostrophic (QG) numerical schemes and GPU-based solvers to simulate flows with coastal boundaries and landforms such as capes.
Then, we benchmark four analytical turbulence closures: the standard Smagorinsky and Leith model, and their dynamic variants, and study their performance in the presence of coasts. Our results highlight the differences between the various models and show the need for the development of new data-driven or analytical closures. 


\section{Quasi-Geostrophic Turbulence and Simulations} \label{sec: QG_highres}

We now describe the fully-resolved (FR) quasi-geostrophic turbulence simulations, including modifications that incorporate coastal boundaries and inflow conditions.

\vspace{-0.25 in}
\subsection{Quasi-Geostrophic Dynamics}

We use the incompressible two-dimensional (2D) Quasi-Geostrophic (QG) equations as a 
common, simple one-layer linearized-Coriolis approximation to oceanic and atmospheric turbulence \cite{majda2006nonlinear,cushman2011introduction}. 
In vorticity form, the non-dimensional QG equations can be written as below, where $\omega$ is the vorticity, our field of interest, $\psi$ is the streamfunction, and $(u,v)$ are the 2D velocity fields. 
\begin{equation} \label{eq:QG}
    \begin{split}
        \frac{\partial \omega}{\partial t} + J(\psi,\omega) & = \frac{1}{Re} \nabla^2 \omega - \mu \omega - \beta \frac{\partial \psi}{\partial x} + F \\
         (u,v) & = (-\frac{\partial \psi }{\partial y}, \frac{\partial \psi }{\partial x}) \\
         \omega & = \nabla^2 \psi  
    \end{split}
\end{equation}
In \eqref{eq:QG}, $J(\psi,\omega)$ is the nonlinear advection term, $Re$ the Reynolds number, $\mu$ the bottom friction coefficient, and $F$ the forcing terms. The $\beta$-term provides a linear approximation to the latitudinal variation of the Coriolis force, leading to the formation of anisotropic zonal jets and striations, prevalent in Jovian atmospheres and the Earth's oceans \cite{galperin2004ubiquitous, xia2020characteristics}. The vorticity form satisfies incompressibility by construction and hence it does not require the computation of pressure \cite{guermond2006overview,aoussou_et_al_JCP2018}.

\subsection{Brinkman Volume Penalization Approach}
We extend the 2D QG equations to flows past coastal boundaries such as islands and interior landforms. One way to numerically simulate these flows is the body-fitted method (BFM), which introduces complex, unstructured grids \cite{bathe2009mesh, ferziger2019computational,deleersnijder_lermusiaux_OD2008} and could require additional affine transformations of the governing equations \eqref{eq:QG}. Another approach is the immersed-boundary method (IBM), which does not modify the grid but instead models a larger, simpler domain and modifies the governing equations in the vicinity of solid boundaries to impose appropriate boundary conditions \cite{peskin2002immersed}. Various modifications to impose these boundary conditions have been studied including the discrete forcing and continuous forcing approaches \cite{mittal2005immersed}. IBM has been shown to be computationally efficient compared to BFM when modeling transient wakes past circular islands and other complex geometries \cite{naderi2024comparison}.

In this work, we utilize IBM with the Brinkman volume penalization approach \cite{arquis1984conditions}. The Brinkman penalty is physically motivated and imposes no-slip boundary conditions by approximating solid obstacles (denoted with subscript $obs$) as porous media with porosity tending to zero. This is achieved by adding a forcing term, $F_{obs}$, to the right hand side of \eqref{eq:QG}, 
\begin{equation} \label{eq:QG_obstacle}
\begin{split}
    F_{obs}(\chi_{obs},\mathbf{u}) &= -  \nabla \times \bigg ( \frac{1}{\eta} \chi_{obs}(x) (\mathbf{u} - \mathbf{u}_{obs}) \bigg) \\   
    \chi_{obs} (x) &= \left\{\begin{array}{@{\mathstrut}l@{\quad}l}
    1 & x \in \Omega_{obs} \\
    0 & x \not \in \Omega_{obs} 
    \end{array}\right.
\end{split}
\end{equation}
where $\nabla \times$ is the curl operator, $\eta$ the porosity, $\chi_{obs}$  the mask function which is $1$ in obstacle cells (land domain $\Omega_{obs}$) and $0$ in fluid cells (ocean), and $\mathbf{u}_{obs}$ the velocity of the obstacle in vector form.
Hence, the flow is governed by the QG equations in the fluid regions, and by Darcy's law in the coastal regions \cite{schneider2005numerical}. For coastal boundaries, we impose the no-slip boundary condition and hence $\mathbf{u}_{obs} = \mathbf{0}$. 

This approach has been utilized to simulate turbulence in tube bundles, complex bounded domains and moving obstacles \cite{schneider2005numerical, schneider2008final, kolomenskiy2009fourier, engels2013two}. Convergence theorems and error estimates for applications to incompressible Navier-Stokes equations with no-slip boundaries have also been studied \cite{angot1999penalization}. However, to our knowledge, it has not yet been used for QG turbulence with coastal boundaries.

\subsection{Spatial Discretization}
To numerically solve the fully-resolved (FR) QG equations, \eqref{eq:QG}, \eqref{eq:QG_obstacle}, we use a Fourier pseudo-spectral spatial discretization on a square domain of size $L \times L$ with $N_{FR} \times N_{FR}$ grid points \cite{canuto2007spectral}. We implement our solver in PyTorch to run on Graphics Processing Units (GPUs), building on \cite{suresh_babu_et_al_2025b, frezat2022posteriori, Lauber_2D-Turbulence-Python_2021}. 
All spatial derivatives are computed in Fourier space, while nonlinear terms such as advection and Brinkman penalty are evaluated by collocation in physical space with full 2/3 de-aliasing to truncate higher-order Fourier coefficients \cite{orszag1974numerical}. The mask function $\chi_{obs}$ is smoothed with a Gaussian filter to avoid Gibbs oscillations \cite{kolomenskiy2010numerical}.
\subsection{Boundary and Inflow Conditions}
Using a Fourier pseudo-spectral spatial discretization assumes that the computational domain is doubly periodic, which is not always valid for oceanic flows. Hence, modifications to handle non-periodic boundary conditions and inflow conditions are required. One way to incorporate non-periodic boundaries is to utilize odd or even extensions to the computational domain using the Discrete Cosine Transform (DCT) instead of the Discrete Fourier Transform (DFT) \cite{strang1999discrete}. An alternate approach, implemented in this work, is to use additional penalization for sponging \cite{engels2013two,haley_lermusiaux_2013prep}. This is achieved by adding another forcing term to the right hand side of \eqref{eq:QG}.
\begin{equation} \label{eq:QG_sponge}
\begin{split}
    F_{sponge}(\chi_{sponge},\omega) &= -  \bigg ( \frac{1}{\eta_{sponge}} \chi_{sponge}(x) \bigg) (\omega - \omega_{sponge})  \\ 
    \chi_{sponge} (x) &= \left\{\begin{array}{@{\mathstrut}l@{\quad}l}
    1 & x \in \partial{\Omega} \\
   1-d(x,y)/d_s & x \in \Omega_s  \\
   0 & x \in \Omega- \Omega_s
    \end{array}\right.
\end{split}
\end{equation}
The sponge mask $\chi_{sponge}$ is here linearly varying 
between 1 and 0 in a thin sponge region $\Omega_s$
along the boundaries $\partial{\Omega}$ of the fluid domain $\Omega$,
where $d(x,y)$ is the distance from the boundary $\partial{\Omega}$ and $d_s$ the sponge thickness.
The sponge term \eqref{eq:QG_sponge} and volume penalization term \eqref{eq:QG_obstacle} have similar convergence properties and hence we set $\eta_{sponge} = \eta$. Applications of sponge terms to open boundary conditions in the ocean and the effects of different sponge functions are studied in \cite{kelly_lermusiaux_JGR2016,lavelle2008pretty,modave2010parameters}.

To apply inflow conditions, a uniform mean flow along the required direction is added to the zeroth Fourier mode of the velocities as a gauge condition \cite{kevlahan2001computation}.
\subsection{Time-stepping Schemes}
For time-stepping, we use a semi-implicit second-order Adams-Bashforth-Crank-Nicolson (AB2CN) scheme, treating the linear terms implicitly (CN) and nonlinear terms explicitly (AB2) for an IMEX scheme \cite{boyd2001chebyshev}.
The explicit treatment leads to a stability constraint on the numerical timestep $\Delta t$ and the porosity parameter $\eta$: $\eta$ must be larger than $\Delta t$ due to numerical stiffness of the volume penalization \cite{menez2023assessment}, so the coastal boundary limit $\eta \to 0$ is not numerically exact. To prevent instabilities arising from overlapping coastal boundaries and sponge layers, we use a splitting scheme \cite{lermusiaux_2.29_notes,mclachlan2002splitting}
where \eqref{eq:QG}--\eqref{eq:QG_obstacle} are solved first using the AB2CN scheme followed by \eqref{eq:QG_sponge} using the AB2 scheme,
\begin{equation}
    \begin{split}
        \omega^{*} & =  \omega^{t} + \frac{\Delta t}{2} \bigg(\frac{1}{Re} \nabla^2 (\omega^{*}+\omega^t) - \mu (\omega^{*}+\omega^t) \\ & - \beta \frac{\partial (\psi^{*}+\psi^t)}{\partial x} \bigg) -  \Delta t \bigg(\frac{3}{2} J(\psi^t,\omega^t) -\frac{1}{2} J(\psi^{t-1},\omega^{t-1})\bigg) \\ & - \nabla \times \bigg (\frac{\Delta t}{\eta} \chi_{obs} \bigg(\frac{3}{2}\mathbf{u}^t - \frac{1}{2}\mathbf{u}^{t-1}\bigg) \bigg) \\
        \omega^{t+1} & = \omega^{*} -  \frac{\Delta t}{\eta} \chi_{sponge} \bigg( \frac{3}{2}\omega^* - \frac{1}{2} \omega^{**} \bigg)
    \end{split}
\end{equation}
where $\omega^{*}$ and  $\omega^{**}$ are the intermediate vorticity fields for timesteps $t \rightarrow t+1$ and $t-1 \rightarrow t$, respectively.

\section{Filtering and Large-Eddy Simulations}\label{sec: QG_lowres}

Next, we describe our filtering operations $\overline{(.)}$, i.e., the grid filter, and large-eddy simulations with subgrid-scale closure. In contrast to the fully-resolved simulations described in Sect.\,\ref{sec: QG_highres}, Large-Eddy Simulations (LES) solve filtered equations where large-scale features are adequately resolved on a coarser grid with $\frac{N_{FR}}{\delta} \times \frac{N_{FR}}{\delta}$ grid points, where $\delta$ is the coarsening scale. Smaller-scale features are approximated through subgrid-scale closure terms \cite{mason1994large}. Various filters have been explored for LES in both physical and spectral space, such as the box filter, Gaussian filter, and spectral cut-off filter \cite{grooms2021diffusion, ross2023benchmarking}. In this work, we first apply the Gaussian and cut-off filters, $\mathcal{H}_{\text{Gaussian}}$ and $\mathcal{H}_{\text{cut-off}}$, in spectral space, then interpolate  the resulting field to the coarse grid with resolution $\Delta_{FF} = \frac{2\pi \times \delta}{N_{FR}}$ in physical space, following \cite{frezat2022posteriori,suresh_babu_et_al_JAMES2025},

\begin{eqnarray}\label{eq:downsample}
     \mathcal{H}_{\text{Gaussian}}(k_{FR}) =  \exp{\bigg(\frac{-k_{FR}^2\;(\delta \times \Delta_{FR})}{6}\bigg)} \\ \nonumber
    \mathcal{H}_{\text{cut-off}}(k) = 0 \; , \qquad \forall\;k > \frac{\pi} {\Delta_{FF}}  \\ \nonumber
    \mathcal{H}_{\text{full}} \triangleq  \text{interpolate}_{\Delta_{FF}}\circ \mathcal{H}_{\text{cut-off}}\circ\mathcal{H}_{\text{Gaussian}} 
\end{eqnarray} 
where $k_{FR}$ is the wavenumber of the fully-resolved grid. 
The Gaussian filter smoothly attenuates high wavenumber content, resembling subgrid-scale energy removal, while the sharp cut-off filter enforces a strict wavenumber limit. This combination enables a robust, stable closure \cite{zhou2019subgrid}.

Consider the final fully-resolved QG equations from Sect.\,\ref{sec: QG_highres},
\begin{equation} \label{eq:QG_FR}
    \begin{split}
        \frac{\partial \omega}{\partial t} + J(\psi,\omega) & = \frac{1}{Re} \nabla^2 \omega - \mu \omega  - \beta \frac{\partial \psi}{\partial x} \\ & +  F_{obs}(\chi_{obs},\mathbf{u}) + F_{sponge}(\chi_{sponge},\omega)  \\
         \omega & = \nabla^2 \psi  
    \end{split}
\end{equation}
Assuming the filters commute with the derivative operators, we obtain the following QG equations for the filtered fields,
\begin{equation} \label{eq:QG_filtered}
    \begin{split}
        \frac{\partial \fvort}{\partial t} + J(\fpsi,\fvort) & = \frac{1}{Re} \nabla^2 \fvort - \mu \fvort  - \beta \frac{\partial \fpsi}{\partial x} \\ & +  F_{obs}(\overline{\chi_{obs}},\fu) + F_{sponge}(\overline{\chi_{sponge}},\fvort) + \Pi \\
        \fvort & = \nabla^2 \fpsi  
    \end{split}
\end{equation}
where $\Pi$ is the \emph{ideal} subgrid-scale (SGS) forcing defined as
\begin{equation} \label{eq:SGS}
    \begin{split}
        \Pi = &   J(\fpsi,\fvort) - \overline{J(\psi,\omega)} \\ & + \overline{F_{obs}(\chi_{obs},\mathbf{u})} - F_{obs}(\overline{\chi_{obs}},\fu) \\ & + \overline{F_{sponge}(\chi_{sponge},\omega)} - F_{sponge}(\overline{\chi_{sponge}},\fvort)
    \end{split}
\end{equation}
The SGS forcing represents momentum and energy exchange between the resolved and unresolved scales, and consists of three components: the first related to the vorticity flux divergence \cite{srinivasan2024turbulence}, the second related to interactions near the obstacle mask arising from the Brinkman penalization term, and the third related to interactions near the sponge. This SGS forcing, $\Pi$, is \emph{ideal} since it contains variables from unresolved scales lost due to filtering. For LES, the SGS forcing needs to be approximated using a closure model that depends only on the resolved variables, $\fvort, \fpsi, \fu$ \cite{pope2001turbulent}.

\subsection{Analytical Subgrid-Scale Closures}

We now describe analytical closures that parametrize $\Pi$ based on the eddy viscosity hypothesis \cite{kraichnan1976eddy}.
In a vorticity-based formulation, unlike momentum-based approaches \cite{srinivasan2024turbulence},
$\Pi$ can be represented directly as follows, where $\nu_e$ is the eddy diffusivity. 
\begin{equation}\label{eq:SGS_eddy}
    \Pi = \nabla.(\nu_e \nabla \fvort)
\end{equation}
Various standard and dynamic formulations of the eddy diffusivity for two-dimensional turbulence have been studied in \cite{maulik2017dynamic, maulik2017stable}. 
We investigate the Smagorinsky, dynamic Smagorinsky, Leith, and dynamic Leith models.

\subsubsection{Smagorinsky Model}
The standard Smagorinsky model \cite{smagorinsky1963general} is one of the earliest approaches for SGS closure, based on the forward cascade of energy in three-dimensional turbulence. Following \cite{maulik2017dynamic}, its eddy diffusivity in the vorticity form is given by
\begin{equation}\label{eq:Smag}
    \begin{split}
        \nu_e = & (c_S \Delta_{FF})^2 \lvert \overline{S} \rvert  \\
        \lvert \overline{S} \rvert = & \sqrt{4\bigg(\frac{\partial^2 \fpsi}{\partial x \partial y}\bigg)^2 + \bigg(\frac{\partial^2 \fpsi}{\partial x^2} -  \frac{\partial^2 \fpsi}{\partial y^2}\bigg)^2} 
    \end{split}
\end{equation}
where $\overline{S}$ is the filtered strain-rate tensor, $\Delta_{FF} = \frac{2 \pi\,\delta}{N_{FR}}$ the filter width (size of the coarse grid), and $c_S$ a parameter to be tuned. In this case, $\nu_e \geq 0$, and hence the closure is purely dissipative. This implies there is no backscatter, i.e, flow of energy from unresolved to resolved scales \cite{pope2001turbulent}.

\subsubsection{Dynamic Smagorinsky Model}
Since the standard Smagorinsky model assumes $c_S$ to be a single universal constant, this model has been observed to over- or under-dissipate if incorrectly tuned \cite{pawar2020priori}. Moreover, $c_S$ is dynamic and spatially varying in turbulent flows near solid boundaries and on rotating frames, thus in geophysical flows with Coriolis \cite{moin1982numerical}. Hence, \cite{germano1991dynamic} proposed a dynamic procedure to estimate $c_S$ using two filters, the grid filter $\overline{(.)}$  with grid resolution $\Delta_{FF}$ \eqref{eq:downsample}, and a test filter, $\widetilde{(.)}$  with resolution typically chosen to be $\widetilde{\Delta}= 2 \times \Delta_{FF}$ \cite{pope2001turbulent}. Using the Germano identity \cite{germano1992turbulence}, the following relationship for $c_S$ can be obtained,
\begin{equation} \label{eq:germano}
    (c_S \widetilde{\Delta})^2 \nabla.(\lvert \widetilde{\overline{S}} \rvert \nabla \widetilde{\fvort}) -  (c_S {\Delta_{FF}})^2 \nabla.(\widetilde{\lvert {\overline{S}} \rvert \nabla {\fvort}})  = J(\widetilde{\fpsi},\widetilde{\fvort}) - \widetilde{{J(\fpsi,\fvort)}}
\end{equation}
Though \eqref{eq:germano} can be solved for $c_S$, it is ill-posed and unstable in regions where $\nabla^2 \fvort \to 0$ \cite{pawar2020priori}. To overcome this instability, \cite{lilly1992proposed} proposed a least-squares estimation technique that provides a global time-varying dynamic parameter,
\begin{equation} \label{eq:lilly_smag}
    \begin{split}
        & (c_S \Delta_{FF})^2  = \frac{\langle H M \rangle}{\langle M^2 \rangle} \\
        & H  = J(\widetilde{\fpsi},\widetilde{\fvort}) - \widetilde{{J(\fpsi,\fvort)}} \\
        & M  = \bigg(\frac{\widetilde{\Delta}}{\Delta_{FF}}\bigg)^2 \nabla.(\lvert \widetilde{\overline{S}} \rvert \nabla \widetilde{\fvort}) -\nabla.(\widetilde{\lvert {\overline{S}} \rvert \nabla{\fvort}}) 
    \end{split}
\end{equation}
where $\langle . \rangle$ denotes spatial averaging and $H$ is related to the residual stress between the grid-filtered field and the test-filtered field (\emph{resolved stress}) \cite{piomelli1993high}. Positive clipping of the numerator, $H$, by setting negative values to zero is also implemented to prevent backscatter \cite{vreman1997large}.

\subsubsection{Leith Model}
Compared to the Smagorinsky model which is based on the forward cascade of energy, the standard Leith model was developed based on the forward cascade of enstrophy \cite{leith1968diffusion}. This hypothesis is more appropriate for two-dimensional turbulence and geophysical flows \cite{leith1971atmospheric}, and has been implemented in mesoscale ocean simulations \cite{fox2008can}. In contrast to the Smagorinsky model which is based on the filtered strain-rate tensor, the Leith model is based on the filtered vorticity gradient $\lvert \nabla \fvort \rvert$,
\begin{equation}\label{eq:Leith}
    \begin{split}
        \nu_e = & (c_L \Delta_{FF})^3 \lvert \nabla \fvort \rvert  \\
         \lvert \nabla \fvort \rvert  = & \sqrt{\bigg(\frac{\partial \fvort}{\partial x}\bigg)^2 +  \bigg(\frac{\partial \fvort}{\partial y}\bigg)^2} 
    \end{split}
\end{equation}
where $c_L$ is a parameter to be tuned. Again in this case, $\nu_e \geq 0$, and there is no backscatter. Modifications to the Leith model to incorporate backscatter have been tested in \cite{grooms2023backscatter}.

\subsubsection{Dynamic Leith Model}
Similar to the Smagorinsky model, a global time-varying dynamic Leith parameter $c_L$ can be obtained using least-squares estimation \cite{lilly1992proposed, maulik2017stable},
\begin{equation} \label{eq:lilly_leith}
    \begin{split}
        & (c_L \Delta_{FF})^3  = \frac{\langle H M \rangle}{\langle M^2 \rangle} \\
        & H  = J(\widetilde{\fpsi},\widetilde{\fvort}) - \widetilde{{J(\fpsi,\fvort)}} \\
        & M  = \bigg(\frac{\widetilde{\Delta}}{\Delta_{FF}}\bigg)^3 \nabla.(\lvert \nabla \widetilde{\fvort} \rvert \nabla \widetilde{\fvort}) -\nabla.(\widetilde{\lvert \nabla \fvort \rvert \nabla {\fvort}}) 
    \end{split}
\end{equation}

\section{Applications}

We now apply the four subgrid-scale closures to idealized quasi-geostrophic turbulent flows past circular islands and capes with the $\beta$-plane approximation. The two dynamic models, Dynamic Smagorinsky \eqref{eq:lilly_smag} and Dynamic Leith $\eqref{eq:lilly_leith}$ compute $c_S$ or $c_L$ in real-time, and only require specification of the test-filter as input. For the standard Smagorinsky and Leith models, we perform a parameter sweep for $c_S$ and $c_L$ and analyze the performance a priori and a posteriori.

We quantify performance with the Pattern \cite{lermusiaux_MWR1999} or Pearson Correlation Coefficient (PCC, Eq.\,\ref{eq:PCC}, as defined in \cite{guan2022stable,zhou2019subgrid}) between the analytical closure ($\Pi_{LES}$), true filtered SGS ($\Pi_{FF}$), and a priori closure ($\Pi_{FR}$) terms.
\begin{equation} \label{eq:PCC}
\begin{split}
    \text{PCC}(X, Y) &= 
    \frac{\bigg\langle(X-\langle X \rangle)\ (Y-\langle Y \rangle)\bigg\rangle}{\sqrt{\bigg\langle(X-\langle X \rangle)^2\bigg\rangle}\sqrt{\bigg\langle(Y-\langle Y \rangle)^2\bigg\rangle}} \\
    \text{PCC}_\text{a-priori} &\triangleq \text{PCC}(\Pi_{FF}, \Pi_{FR}) \\
    \text{PCC}_\text{a-post. SGS} &\triangleq \text{PCC}(\Pi_{LES}, \Pi_{FF}) \\
    \text{PCC}_{\text{a-post. }\omega} &\triangleq\text{PCC}(\omega_{LES}, \omega_{FF}) \\
\end{split}
\end{equation}

\subsection{Flow Past Circular Islands}

\begin{figure}[]
    \centering
    \includegraphics[width=0.75\linewidth]{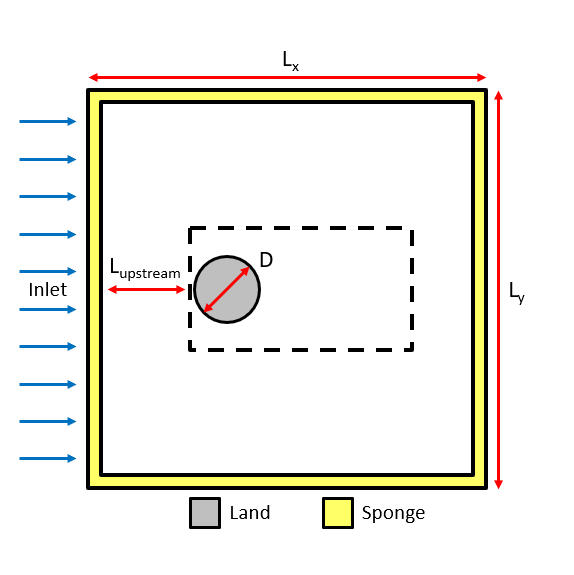}
    \caption{ \small Set-up of domain for flow past a cylinder or idealized island. The computational domain consists of land (grey) and a sponge layer (yellow). Dotted black lines show the region of interest.}
    \label{fig:fpc_domain}
\end{figure}

Our first application is for flows past a cylinder or idealized island with Reynolds number $\text{Re}=200$ for mild turbulence. It is used in part for method checks and software implementation validations. In this case, we use $\beta=0$, i.e, an $f$-plane approximation. Fig.\;\ref{fig:fpc_domain} shows the set-up of the domain along with the land masks, sponge, and region of interest. For all simulations, we use porosity $\eta = 1.025\,dt$. Following \cite{graham2013framework}, we use a length-scale of $\frac{126}{\pi} \times 10^4$ m and time-scale of $1.2 \times 10^6$ s, corresponding to ocean mesoscales. In non-dimensional units, we use $L_x = L_y = 8 \pi$, $D = \frac{2 \pi}{5}$, $L_{upstream} = 0.35 L_x$, with sponge of width $0.05 L_x$ along all four boundaries. The inlet velocity is set to be constant at $2$.

\begin{figure*}[h]
    \centering
    \subfloat[][Evolution of a posteriori SGS correlation \\ $(\text{PCC}_\text{a-post. SGS})$ of $\Pi$]
    {\includegraphics[width=0.32\linewidth]{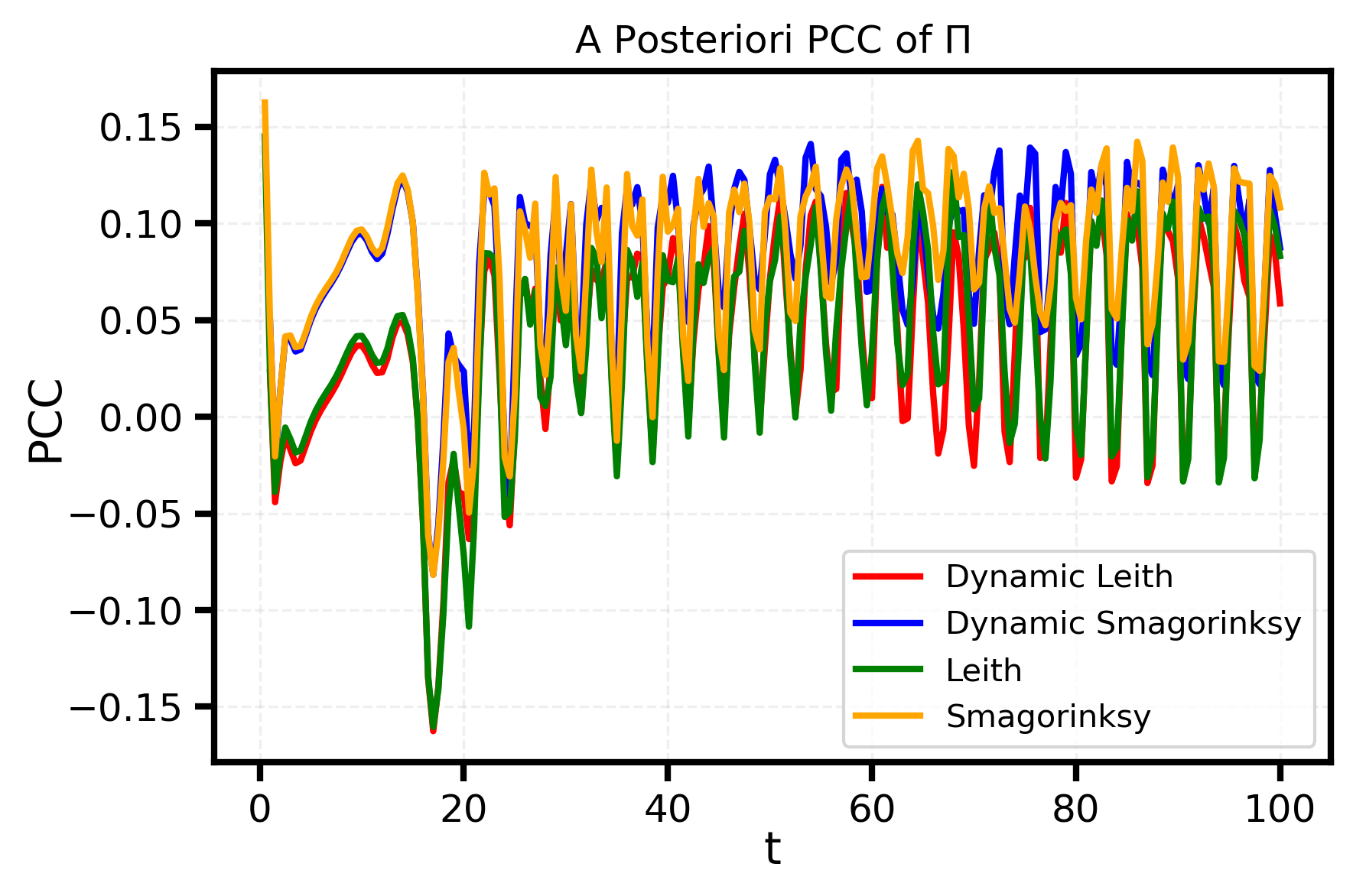}~}
    ~\;\subfloat[][\small Evolution of the eddy diffusivity \\ coefficient $c$]
    {\includegraphics[width=0.32\linewidth]{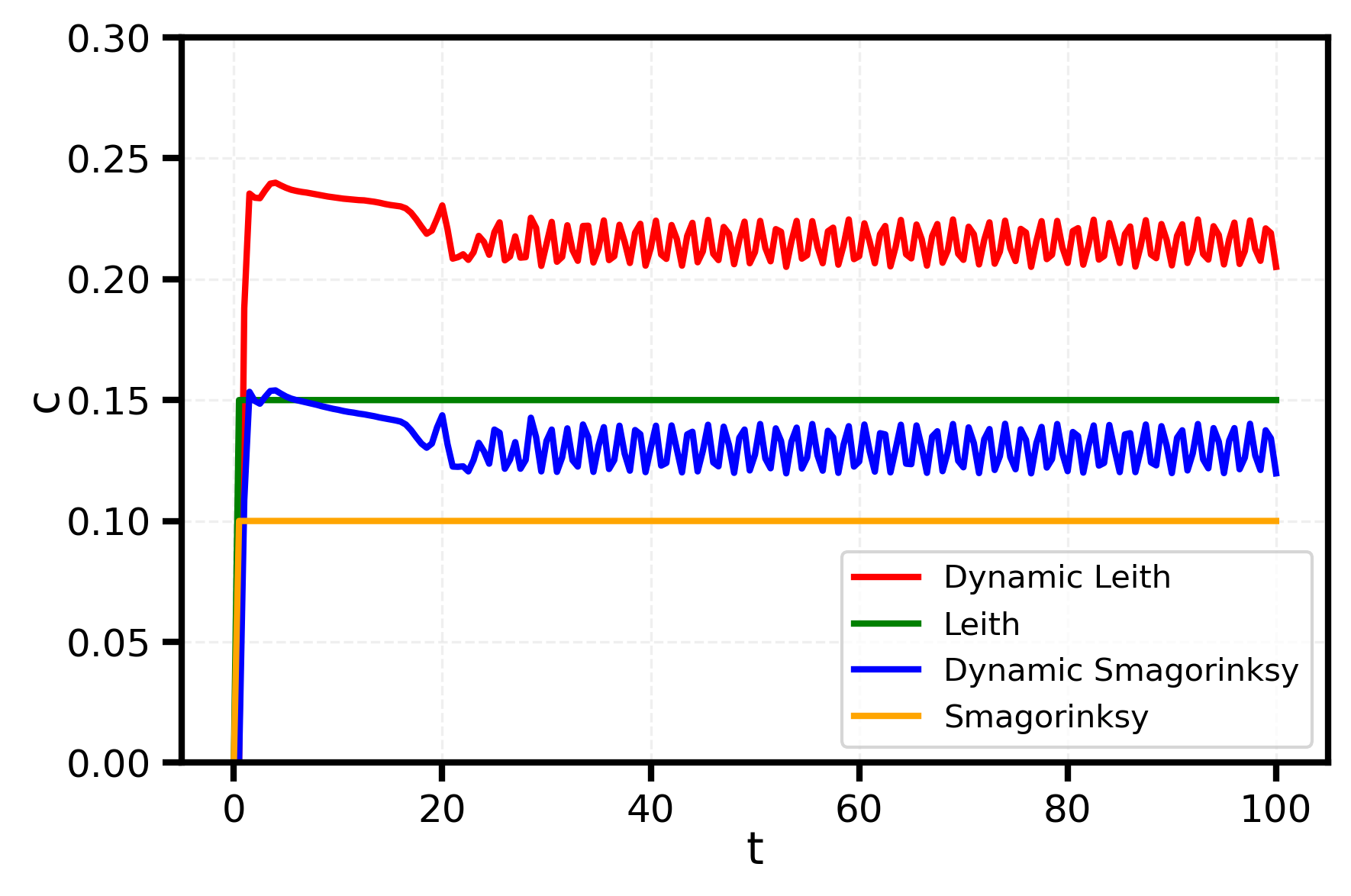}}
    ~~\;\subfloat[][\small Evolution of a posteriori correlation $(\text{PCC}_{\text{a-post. }\omega})$ of $\omega$]
    {\includegraphics[width=0.32\linewidth]{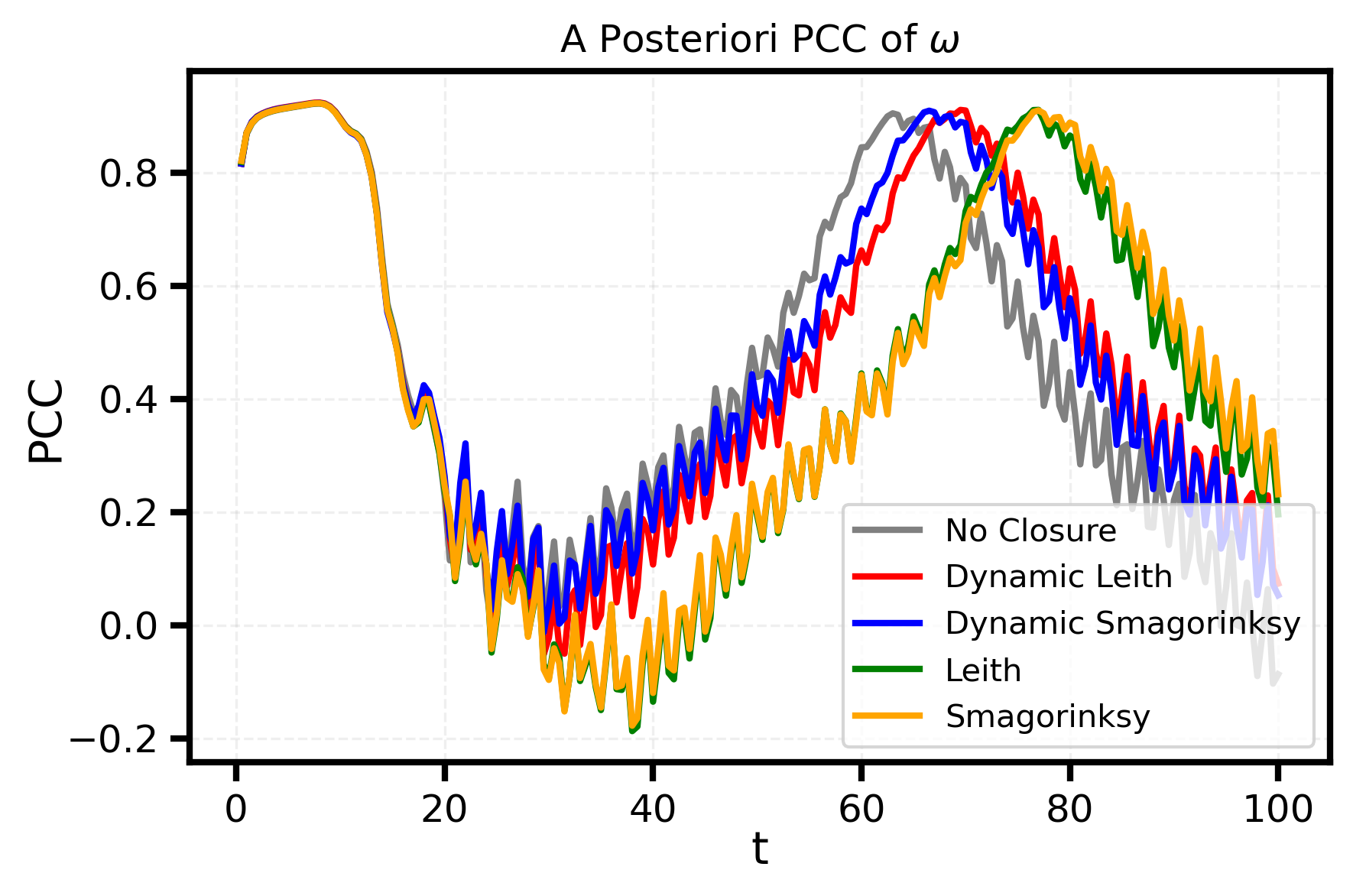}}
    \caption{ \small Evolution of parameters and metrics over time for flow past circular islands with $\beta=0$}
    \label{fig:metrics_fpc}
\end{figure*}

\begin{figure*}[hb]
    \centering
    \subfloat[][Filtered Field]
    {\includegraphics[width=0.3\linewidth]{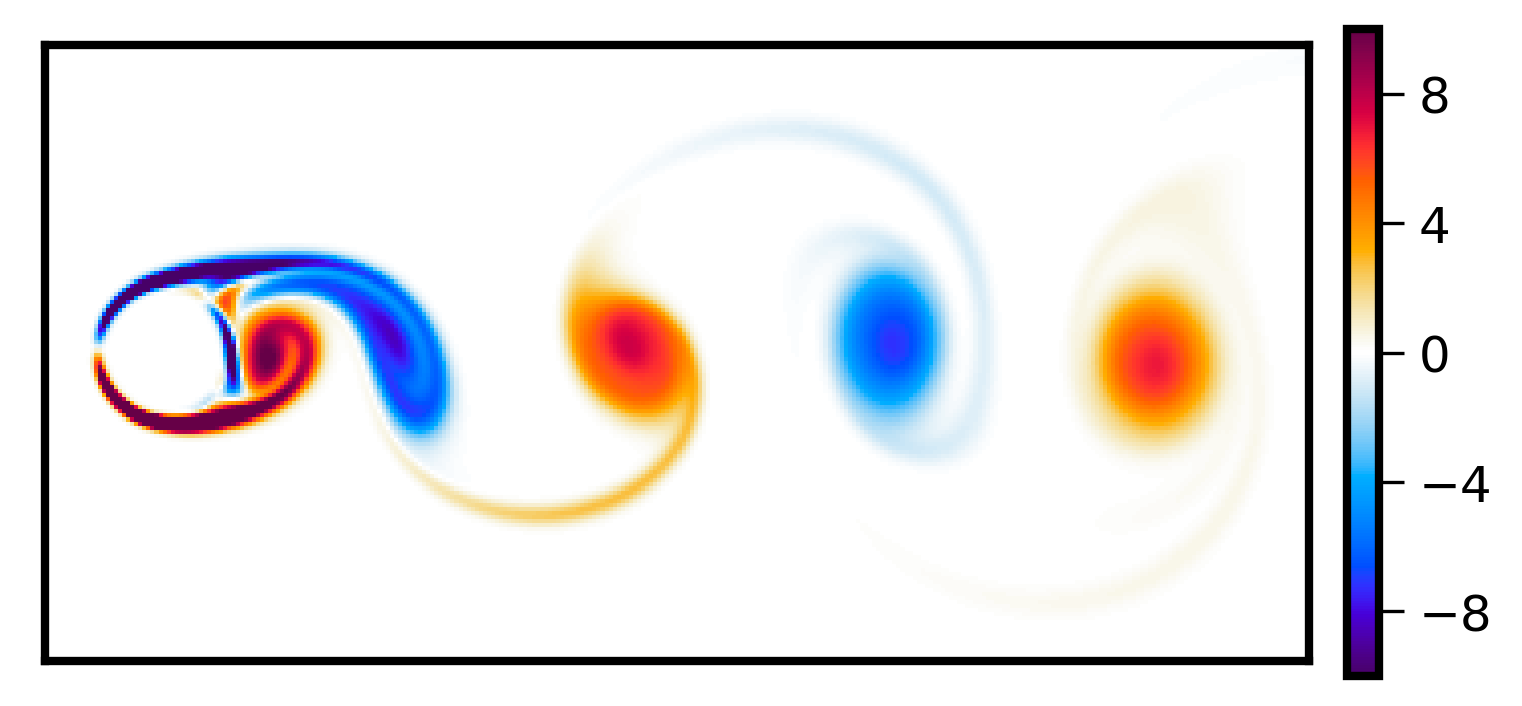}}
    \subfloat[][No Closure]
    {\includegraphics[width=0.3\linewidth]{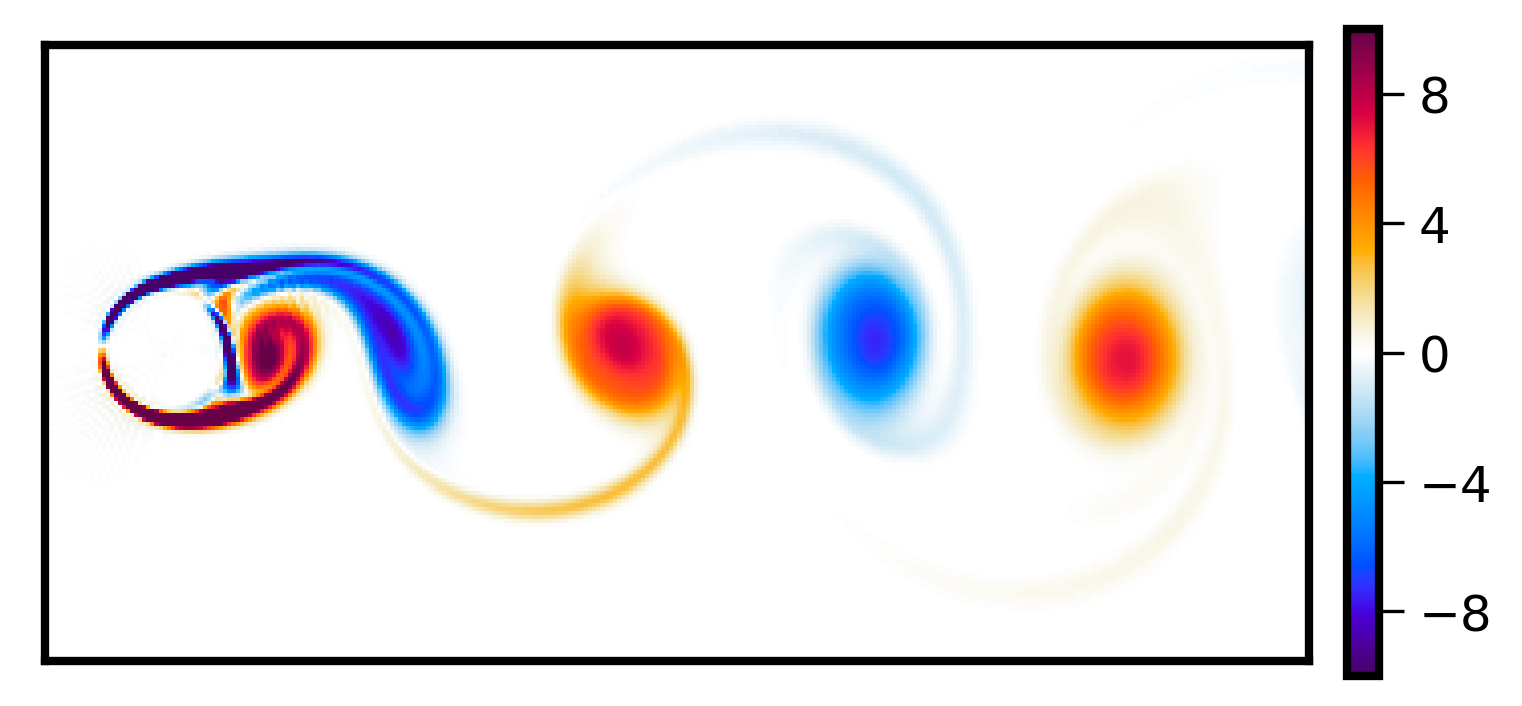}}
    \subfloat[][Dynamic Smagorinsky]
    {\includegraphics[width=0.3\linewidth]{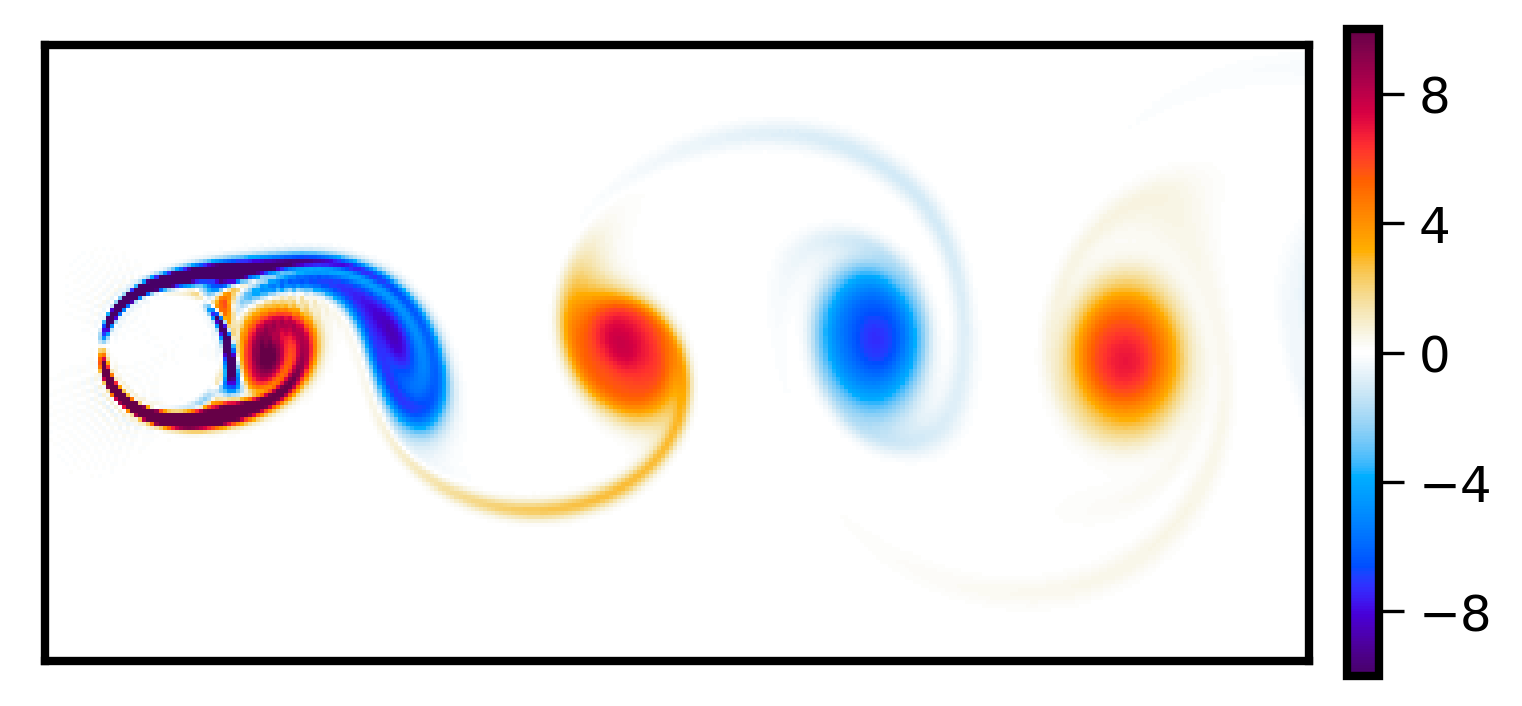}} \\
    \subfloat[][Dynamic Leith]
    {\includegraphics[width=0.3\linewidth]{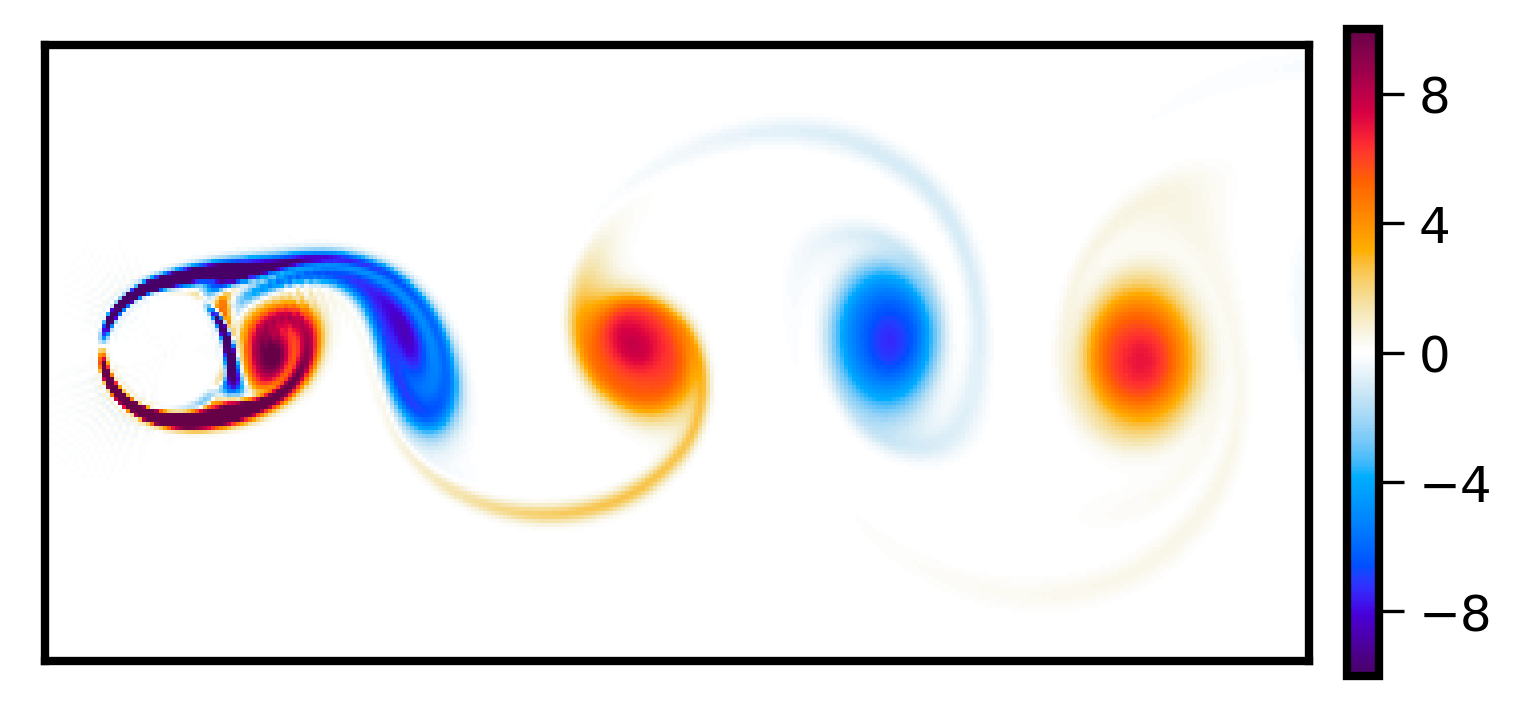}}
    \subfloat[][Smagorinsky]
    {\includegraphics[width=0.3\linewidth]{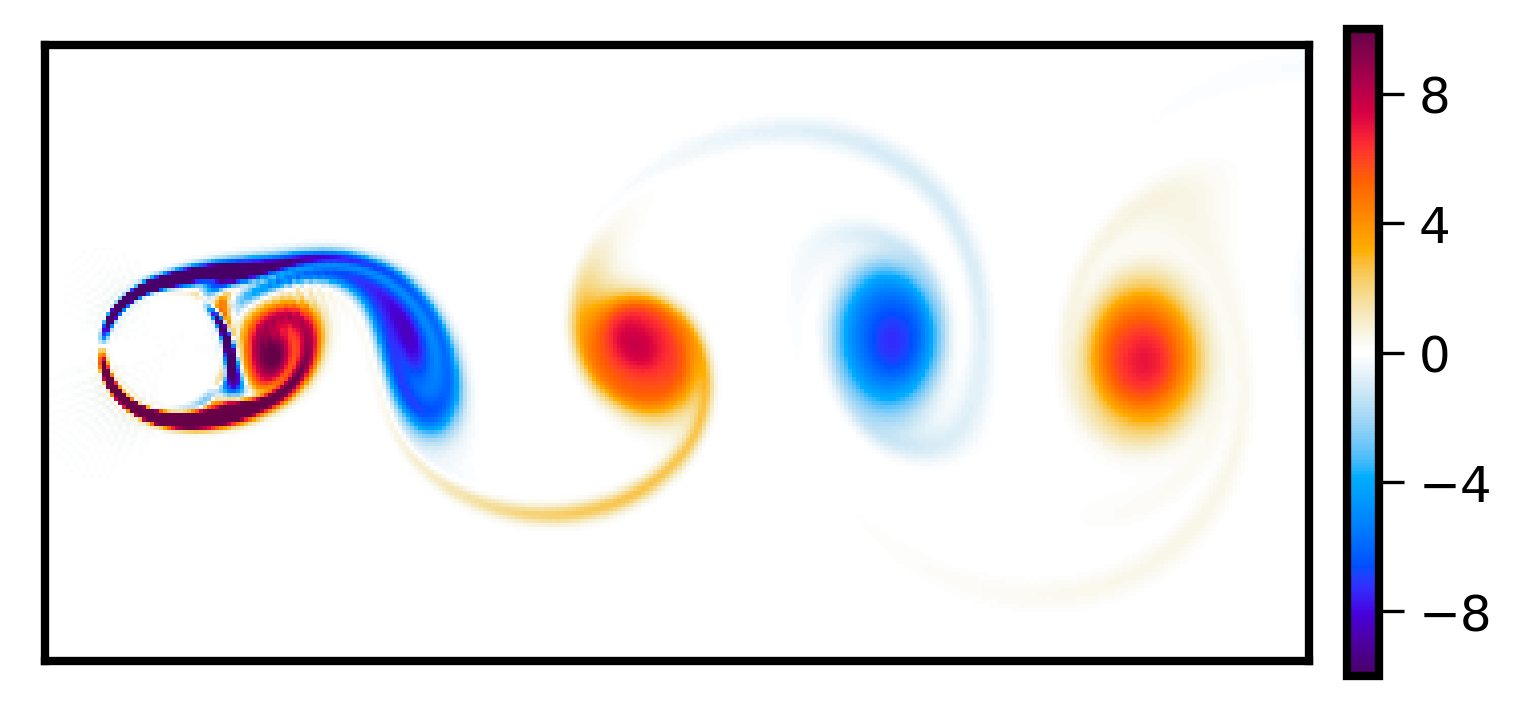}}
    \subfloat[][Leith]
    {\includegraphics[width=0.3\linewidth]{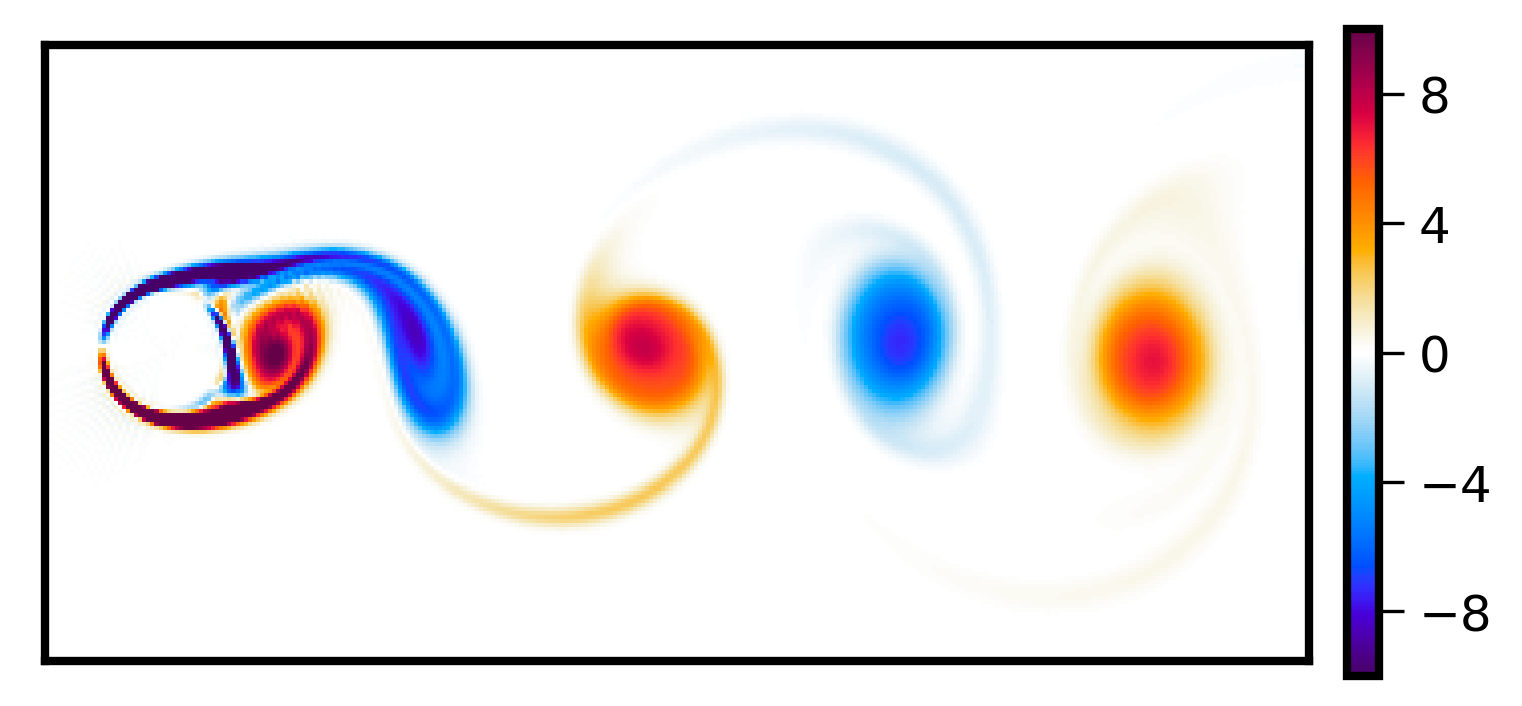}}
    \caption{ \small Vorticity plot of the filtered field (a), and vorticity fields of LES most similar (minimum phase shift error) to the filtered field (b-f) for flows past the circular island with $\beta=0$}
    \label{fig:field_fpc}
\end{figure*}

As a reference solution, we use a fully-resolved (FR) simulation with grid size $N_{FR} \times N_{FR} = 1024 \times 1024$, and obtain the filtered field (FF) with coarsening scale $\delta=2$. The FR simulation is run for a total time of $T = 100$ with $dt = 10^{-4}$. Snapshots are saved every 5000 timesteps. LES simulations with and without closure are simulated with a coarsened grid of size $512 \times 512$ ($\delta=2$), with $dt = 10^{-3}$, i.e., $10 \times$ larger than that of FR. Snapshots are saved every 500 timesteps for the LES. We also studied the performance of closures at filtered resolution of $256 \times 256$, and observed similar trends in the results (figures not shown).

\subsubsection{A Priori Analysis}
We first consider an a priori analysis, where only the estimation of the SGS term $\Pi$ \eqref{eq:SGS} by the closure model is evaluated using the FR fields and FF, as opposed to analysis of the simulated LES fields \cite{pope2001turbulent}. 

Table\,\ref{tab:FPC_priori_PCC} shows the time-averaged a priori PCC ($\text{PCC}_\text{a-priori}$) for the four analytical closures. We note that the values of $c$ do not affect the a priori analysis, as the analysis only depends on the FR fields and FF. The Leith model performs slightly better than the Smagorinsky model, although both models show moderate-to-weak correlation.
\begin{table}[]
    \centering
    \begin{tabular}{c|c|c}
          \hline
          & (Dynamic) Smagorinsky & (Dynamic) Leith\\
         \hline
         PCC & 0.3516 & 0.4312\\
        \hline
    \end{tabular}
    \caption{ \small Time-averaged a priori correlation $(\text{PCC}_\text{a-priori})$ of $\Pi$ for flow past circular islands with $\beta=0$}
    \label{tab:FPC_priori_PCC}
\end{table}

\subsubsection{A Posteriori Assessment}
Fig.\;\ref{fig:metrics_fpc}(a) shows SGS PCC ($\text{PCC}_\text{a-post. SGS}$) for the four analytical closures with optimal\;$c$. On average, all four closure models exhibit relatively weak performance, with the Smagorinsky closures performing twice as well as the Leith closure.
Fig.\;\ref{fig:metrics_fpc}(b) shows the evolution of the eddy diffusivity coefficient $c$ over time for the dynamic and standard variants (where $c$ was chosen to minimize $\text{PCC}_\text{a-post. SGS}$). Obtained values are close to flow-past-cylinder literature \cite{murakami1997analysis, tian2020new}. As observed in \cite{maulik2017dynamic}, the dynamic values of $c$ increase to their maximum value at the beginning and gradually decay to their steady state values. After spin-up of the simulation (0-18 time units), vortex shedding causes oscillations with frequency close to the Strouhal number. The steady state of the dynamic closures varies from the optimal values of the standard variants, and the larger discrepancy for Leith affirms the better Smagorinsky performance.

As with the dynamic evolution of $c$, the SGS PCC also oscillates post spin-up with frequency close to the Strouhal number. Both dynamic closures have higher peak-to-peak variation, but perform slightly worse on average compared to the standard variants.
We now analyze vorticity. While SGS metrics provide a performance estimate of eddy diffusivity-based closures, their vorticity performance can be different \cite{guan2022stable}. 
In this application, the values of $c_S$ and $c_L$ that optimize for SGS remain optimal for vorticity.
We first study the a posteriori correlation ($\text{PCC}_{\text{a-post. }\omega}$) of the simulated vorticity fields \eqref{eq:PCC} and show their temporal evolution in Fig.\;\ref{fig:metrics_fpc}(c).

\begin{table}[]
    \centering
    \begin{tabular}{c|c|c|c|c}
          \hline
          No Closure & Dyn. Smag. & Dyn. Leith & Smag. & Leith   \\
          \hline
        5.5 & -1 & 2.5 & -0.5 & -0.5 \\
        \hline
    \end{tabular}
    \caption{ \small Phase shift (in non-dimensional time units) between the filtered field and LES for flow past circular islands with $\beta=0$}
    \label{tab:phase_fpc}
\end{table}

During spin-up, none of the closures show improvement, and the effect of shedding oscillation is again visible. The near-identical, phase-shifted PCC profiles in a priori (figures not shown) and a posteriori imply limited spatial feature modifications by the closures. We quantify these phase shifts $\phi$ in \ref{eq:phase}.
\begin{equation} \label{eq:phase}
    \phi = \underset{\phi}{\text{argmin}} \lVert \omega_\text{LES}({t-\phi}) -\omega_\text{FF}(t) \rVert_2
\end{equation}
Table\,\ref{tab:phase_fpc} indicates that the optimal (non-dynamic) Smagorinsky and Leith closures are more in-phase with the filtered, fully-resolved (FF) fields, while the dynamic closures are more in phase with the unclosed LES. This suggests that while the dynamic closures appear more accurate in Fig.\;\ref{fig:metrics_fpc}(c), over a longer predictive horizon, standard closures could be more effective. We emphasize that phase error is the primary source of error: in Fig.\;\ref{fig:field_fpc}, the phase-corrected plots are close to indistinguishable. We also analyzed the power spectra of the vorticity fields, but do not show them here as they are quite close and any variation is overpowered by vortex shedding oscillation.

\begin{figure}[h]
    \centering
    \includegraphics[width=0.75\linewidth]{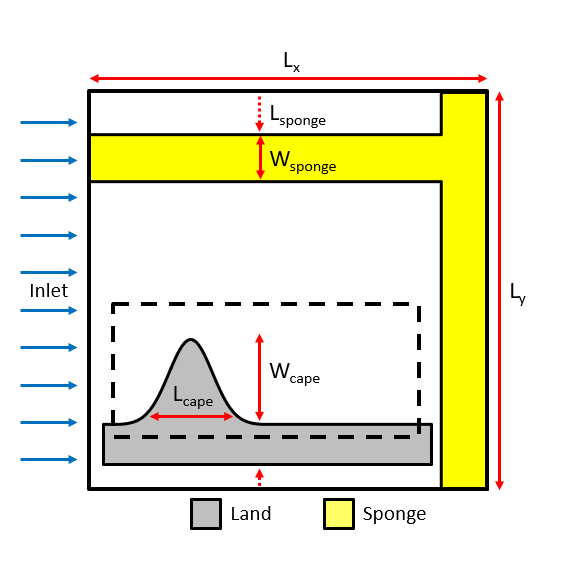}
    \caption{ \small Set-up of domain for flow past an idealized cape. The computational domain consists of land (grey) and a sponge layer (yellow). Dotted black lines show the region of interest.}
    \label{fig:cape_domain}
\end{figure}

\begin{figure}[h]
    \centering
    \subfloat[][$\beta = 0$]
    {\includegraphics[width=0.55\linewidth]{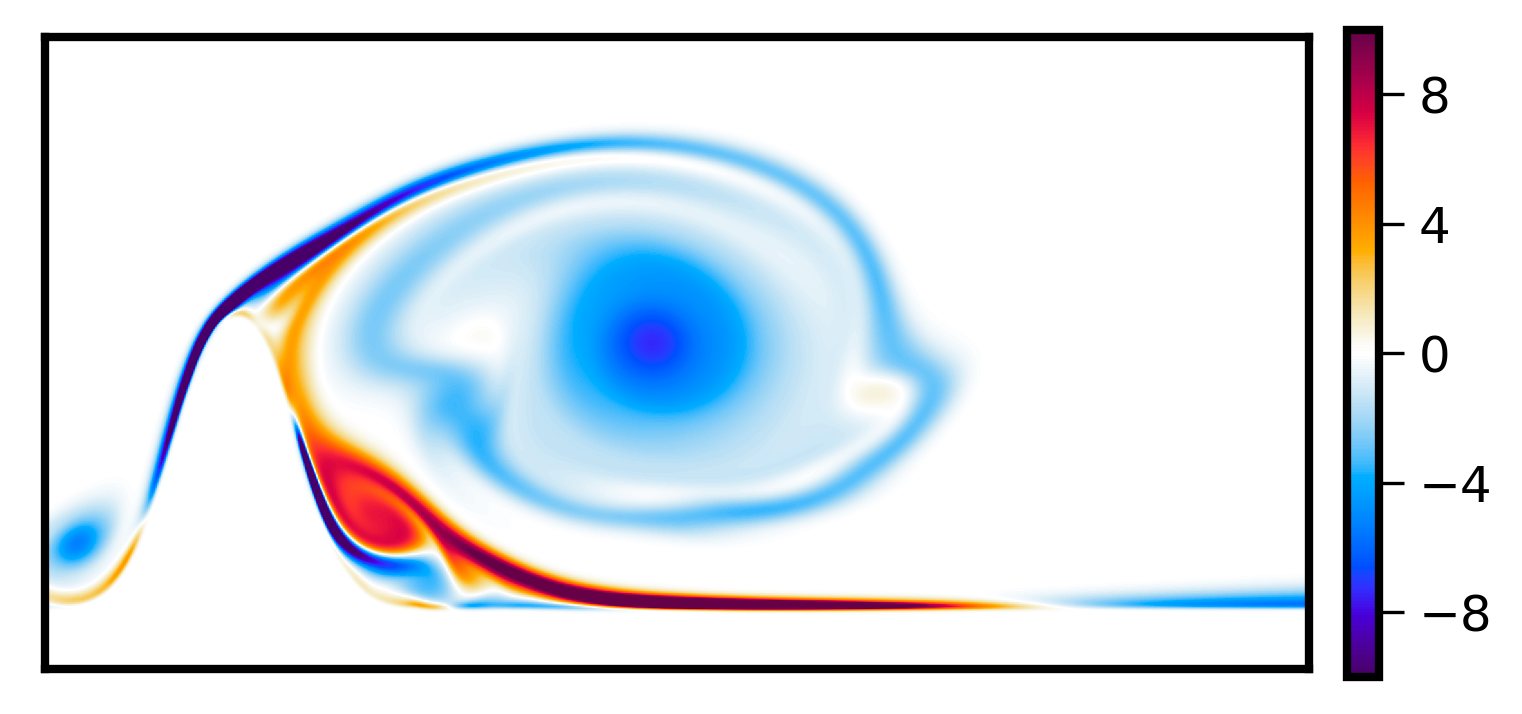}}
    \subfloat[][$\beta = 0.1$]
    {\includegraphics[width=0.55\linewidth]{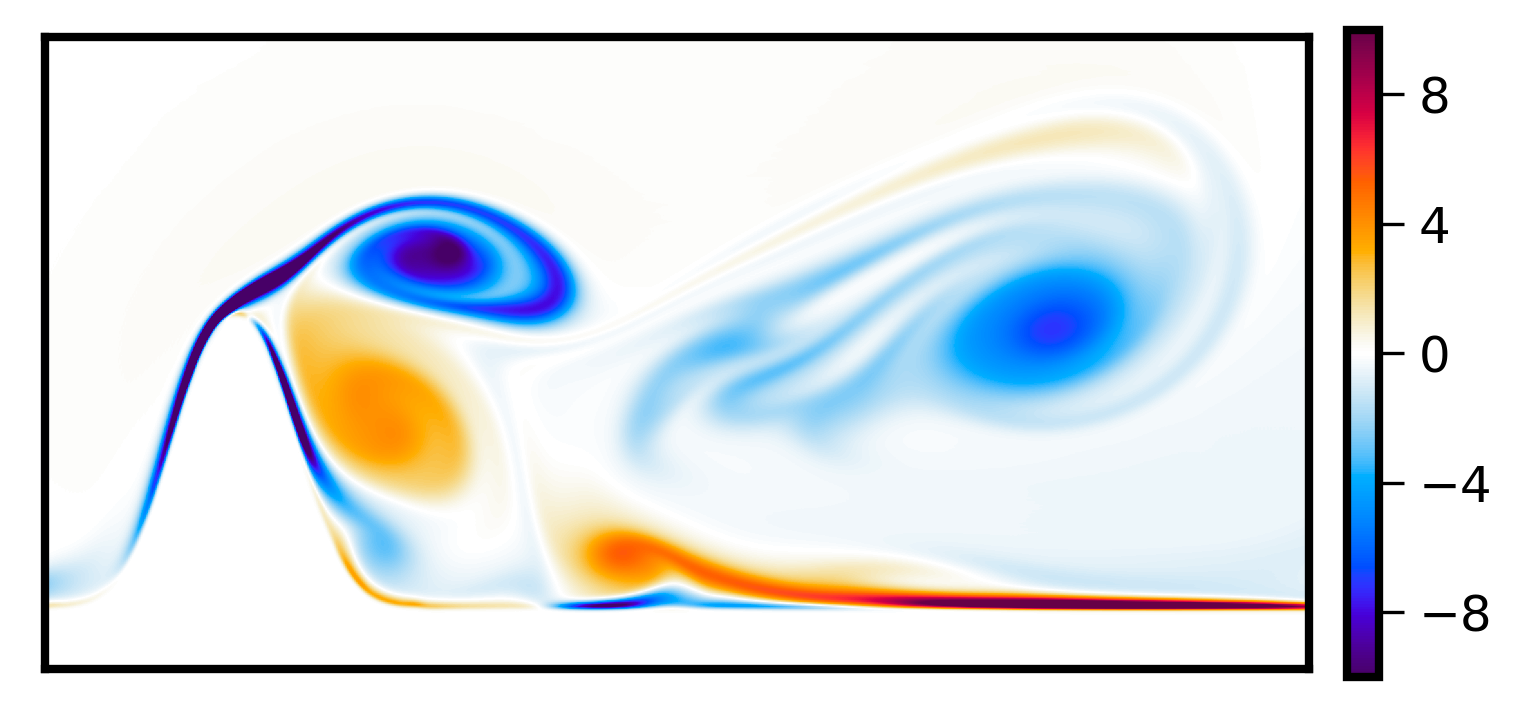}} \\
    \subfloat[][$\beta = 1$]
    {\includegraphics[width=0.55\linewidth]{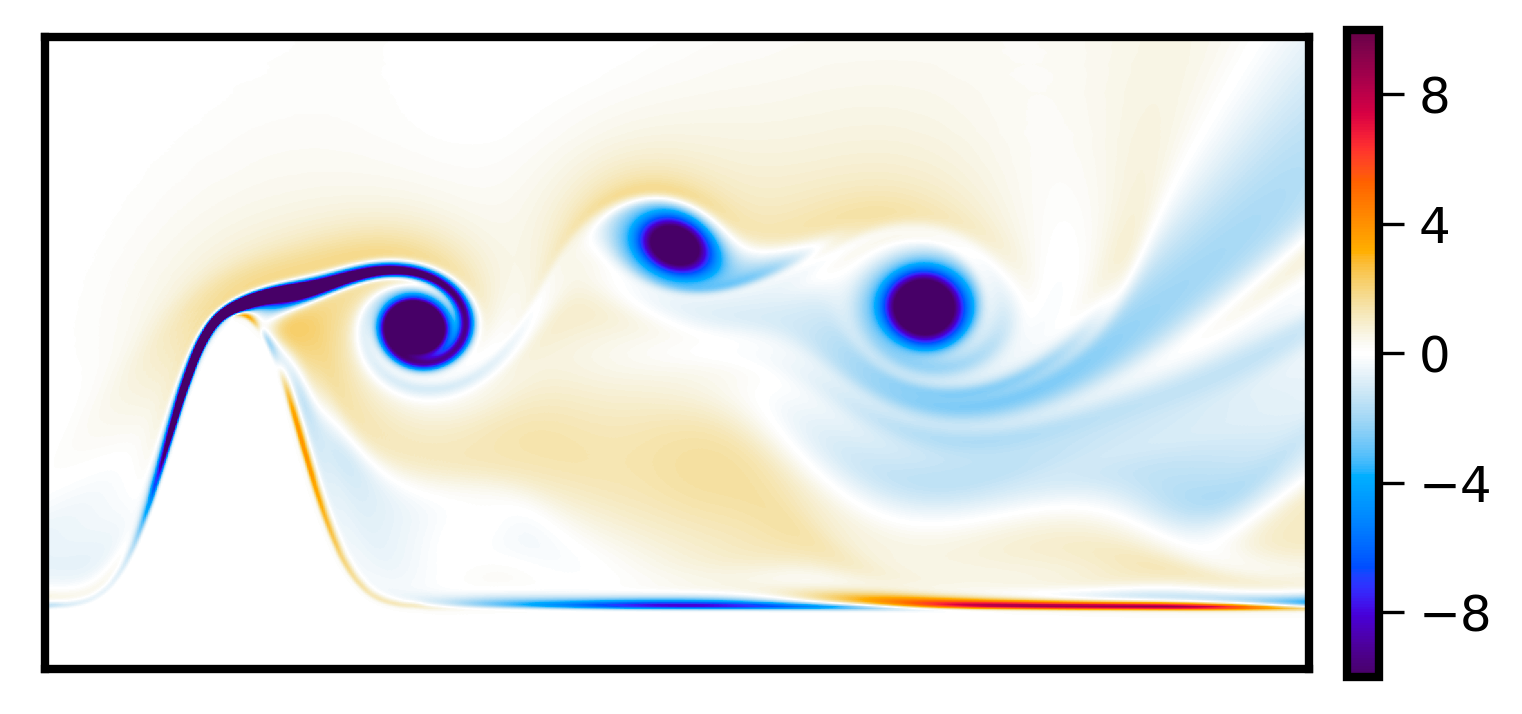}}
    \caption{\small Fully-resolved vorticity fields for flows past an idealized cape with various $\beta$, all at the same time instant.}
    \label{fig:DNS_cape}
\end{figure}

\subsection{Flow Past Idealized Capes}

We now consider flows past idealized capes, once again at $\text{Re}=200$. In this application, we also study the effect of the latitudinal-variation of Coriolis ($\beta$). Previous studies have focused on simulating QG flows past capes and coastal headlands \cite{verron1991quasigeostrophic, davies1995eddy}, although closure of these flows has not yet been studied. Fig.\;\ref{fig:cape_domain} shows the set-up of the domain along with the land masks, sponge and region of interest.  
In non-dimensional units, we again use $L_x = L_y = 8 \pi$. The inlet velocity is set to be constant at $2$. The idealized cape is defined as a Gaussian function, $y \leq W_{\text{cape}}\exp(-[(x - x_{\text{center}})/L_{\text{cape}}]^2)$, with values $L_{\text{cape}}=1$, $W_{\text{cape}}=4$, $x_{\text{center}}=0.2\times L_x$ and sponge parameters $L_{\text{sponge}}=0.125 \times L_x$, $W_{\text{sponge}}=0.1 \times L_x$.

\begin{figure*}[h]
    \centering
    \subfloat[][$\beta = 0$]
    {\includegraphics[width=0.35\linewidth]{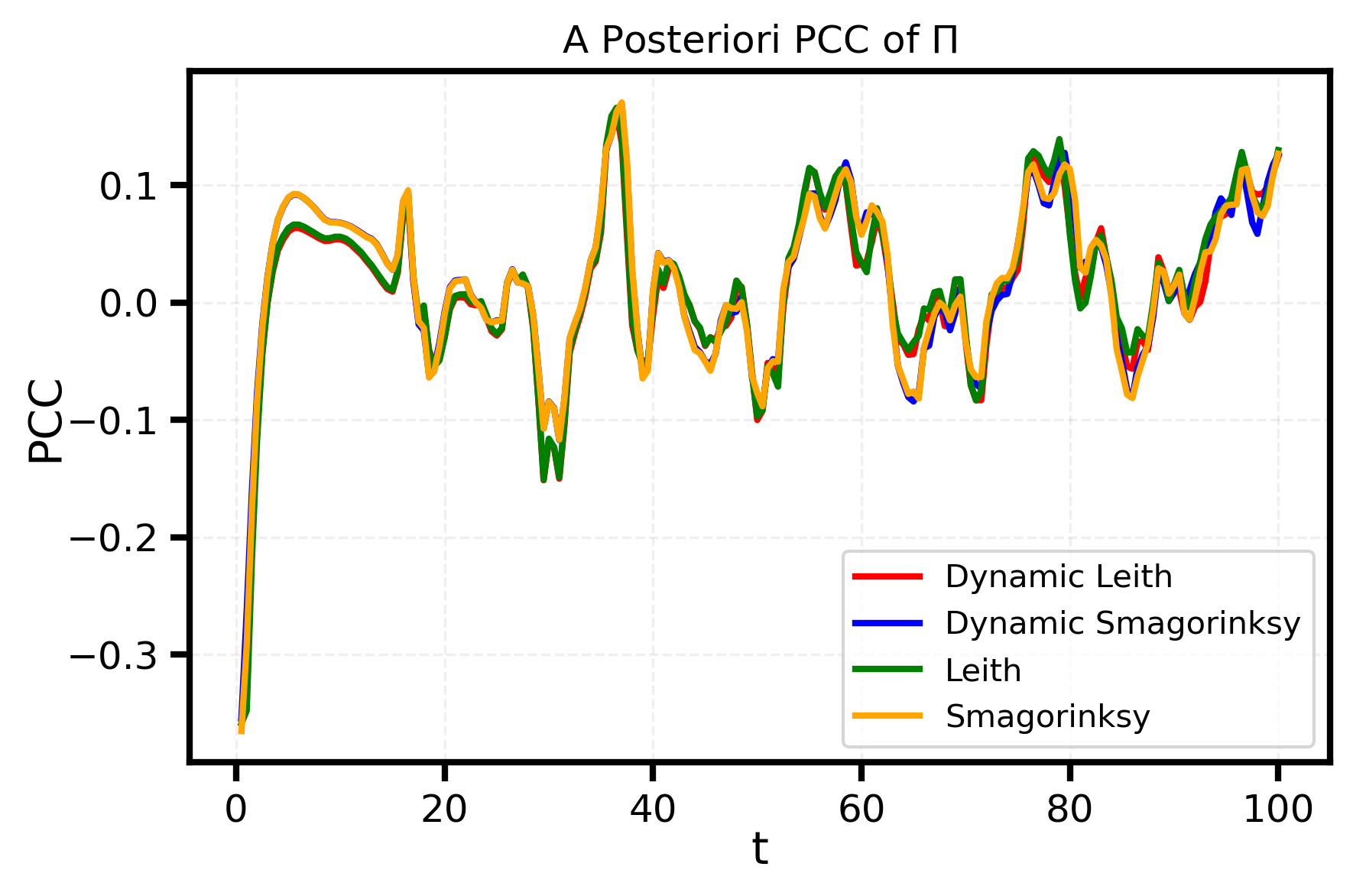}}
    \subfloat[][$\beta = 0.1$]
    {\includegraphics[width=0.35\linewidth]{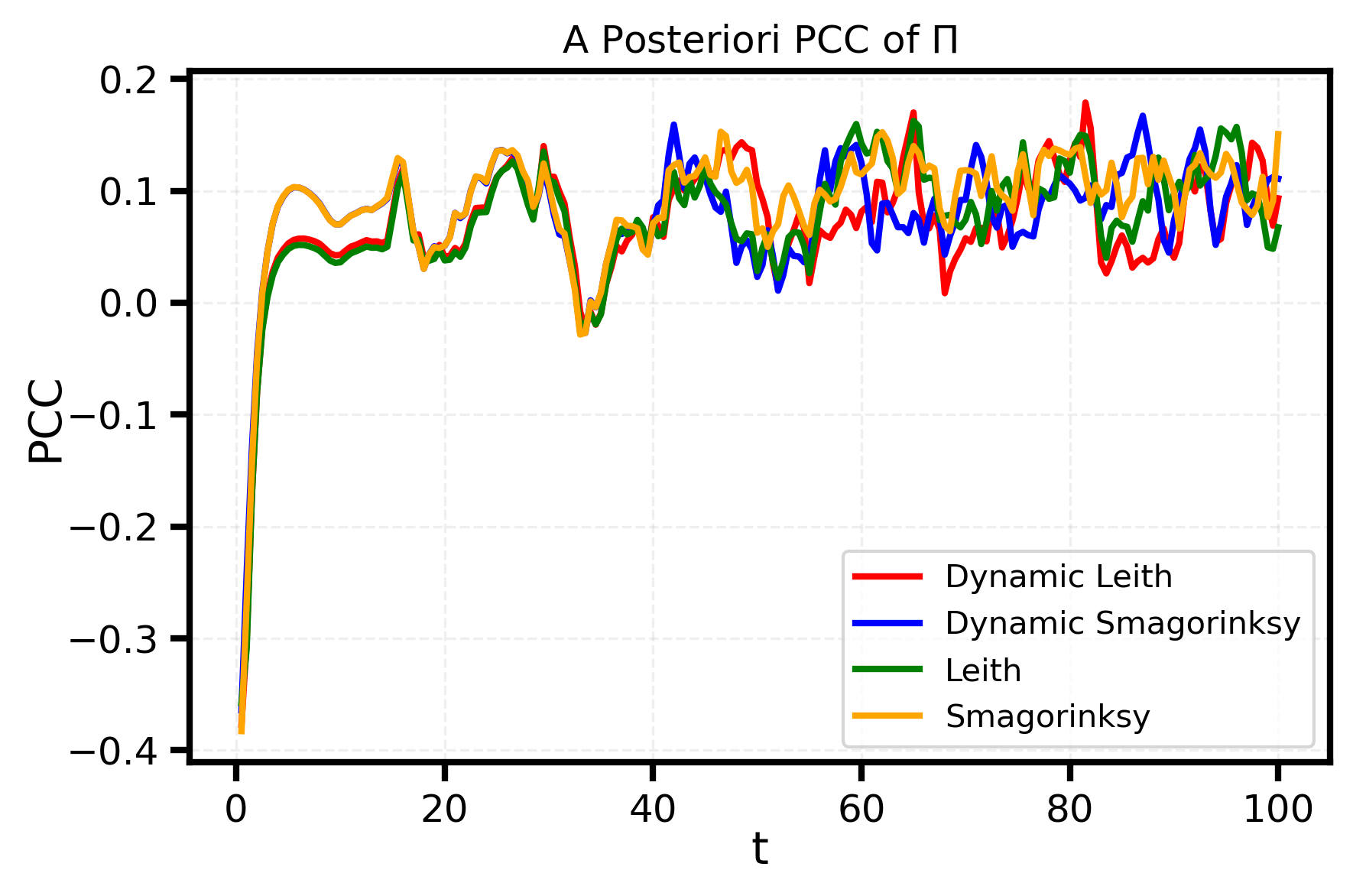}}
    \subfloat[][$\beta = 1$]
    {\includegraphics[width=0.35\linewidth]{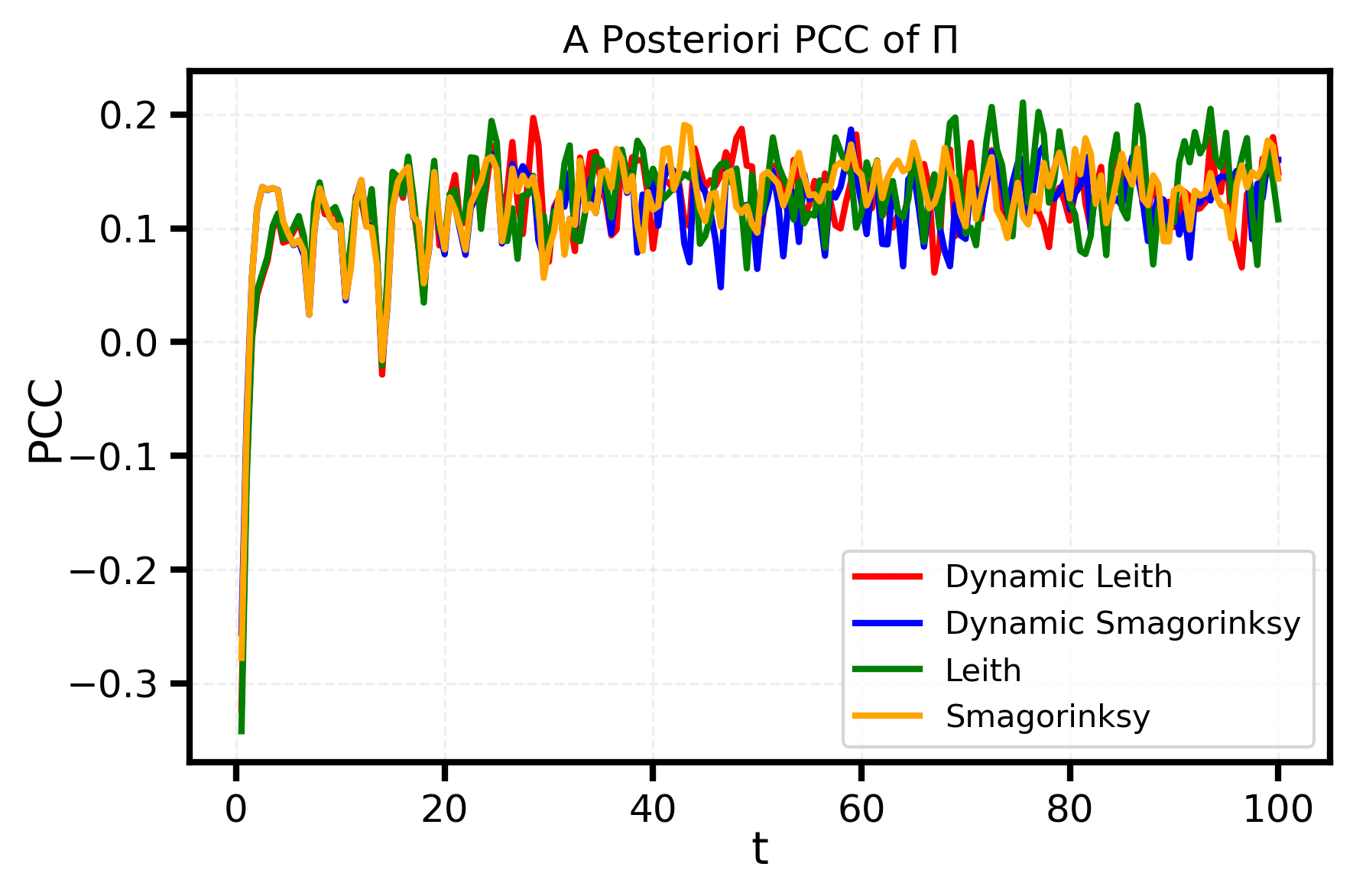}}
    \caption{ \small Evolution of a posteriori SGS correlation $(\text{PCC}_\text{a-post. SGS})$ of $\Pi$ for flows past an idealized cape with various $\beta$}
    \label{fig:SGS_cape}
\end{figure*}

\begin{figure*}[h]
    \centering
    \subfloat[][$\beta = 0$]
    {\includegraphics[width=0.35\linewidth]{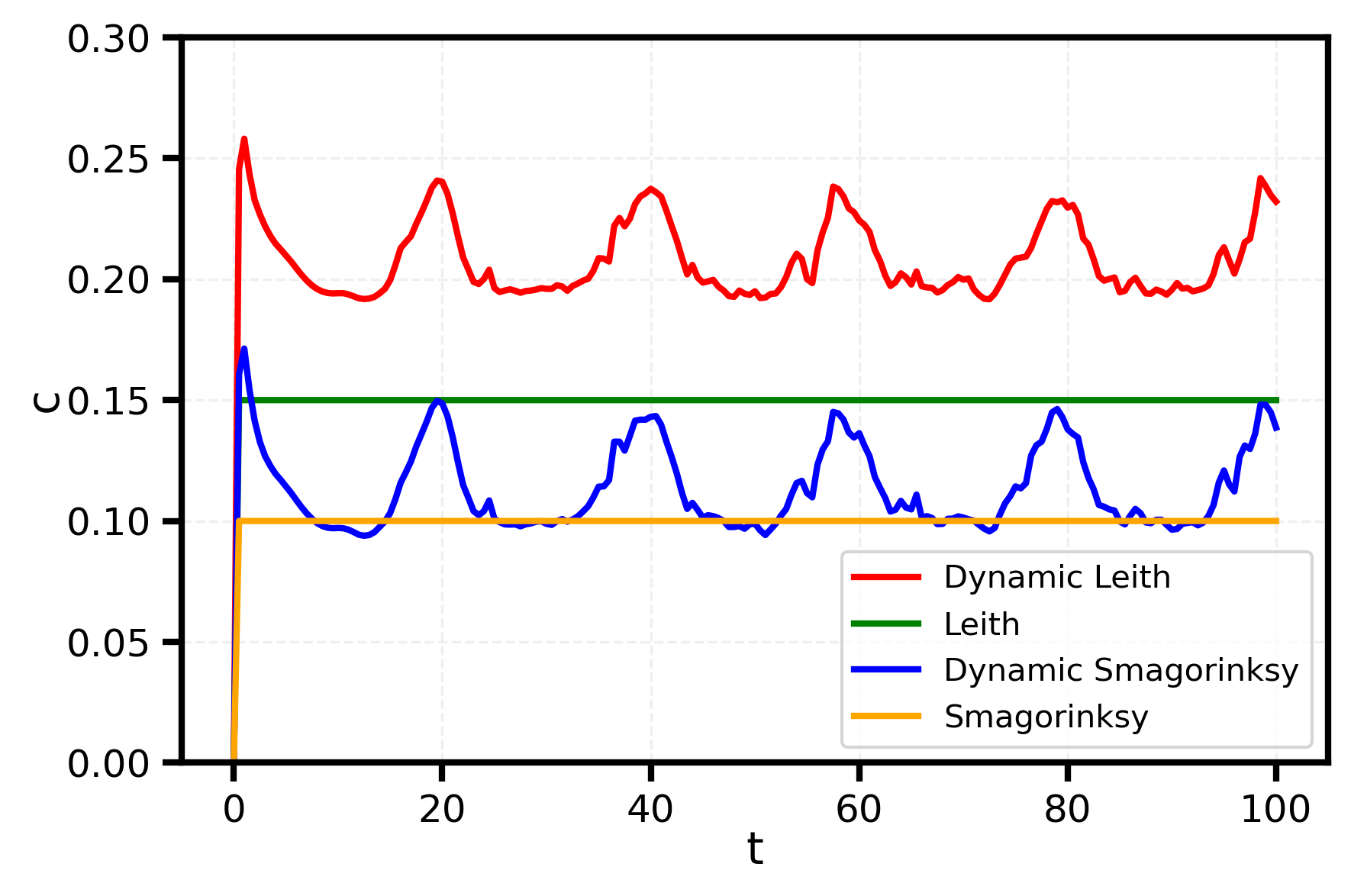}}
    \subfloat[][$\beta = 0.1$]
    {\includegraphics[width=0.35\linewidth]{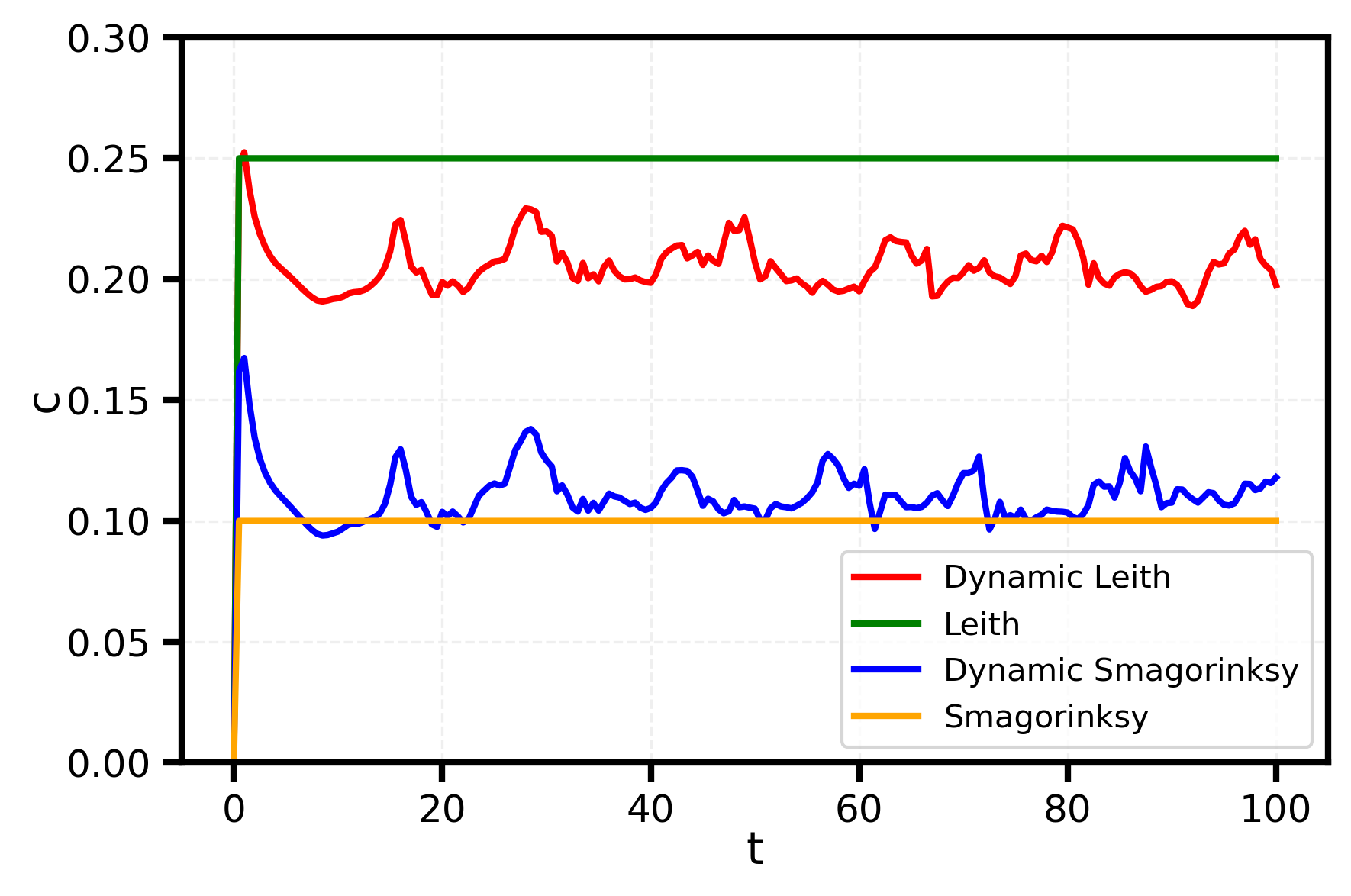}}
    \subfloat[][$\beta = 1$]
    {\includegraphics[width=0.35\linewidth]{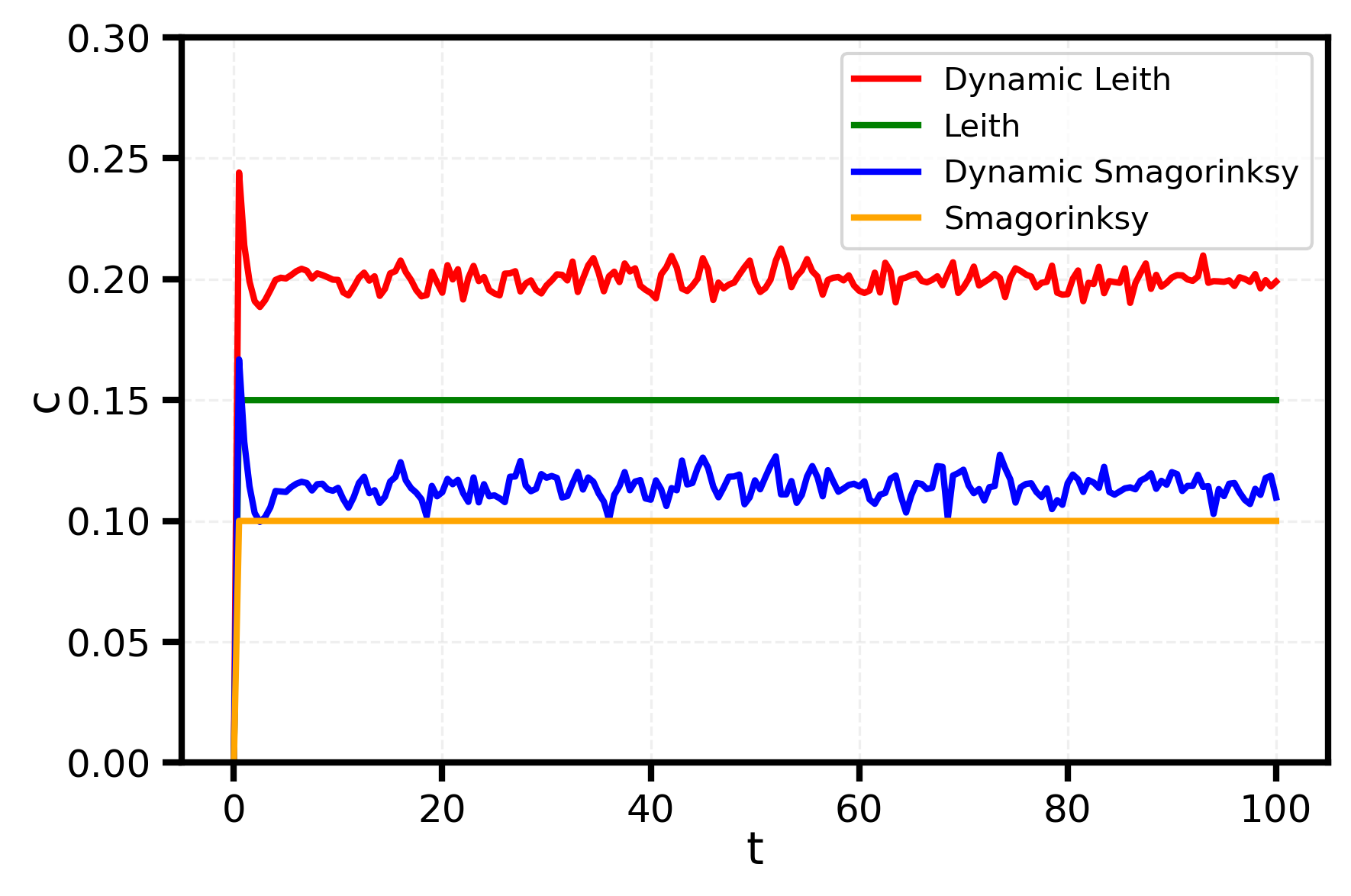}}
    \caption{ \small Evolution of the eddy diffusivity coefficient $c$ over time for flows past an idealized cape with various $\beta$}
    \label{fig:c_cape}
\end{figure*}

\begin{figure*}[h]
    \centering
    \subfloat[][$\beta = 0$]
    {\includegraphics[width=0.35\linewidth]{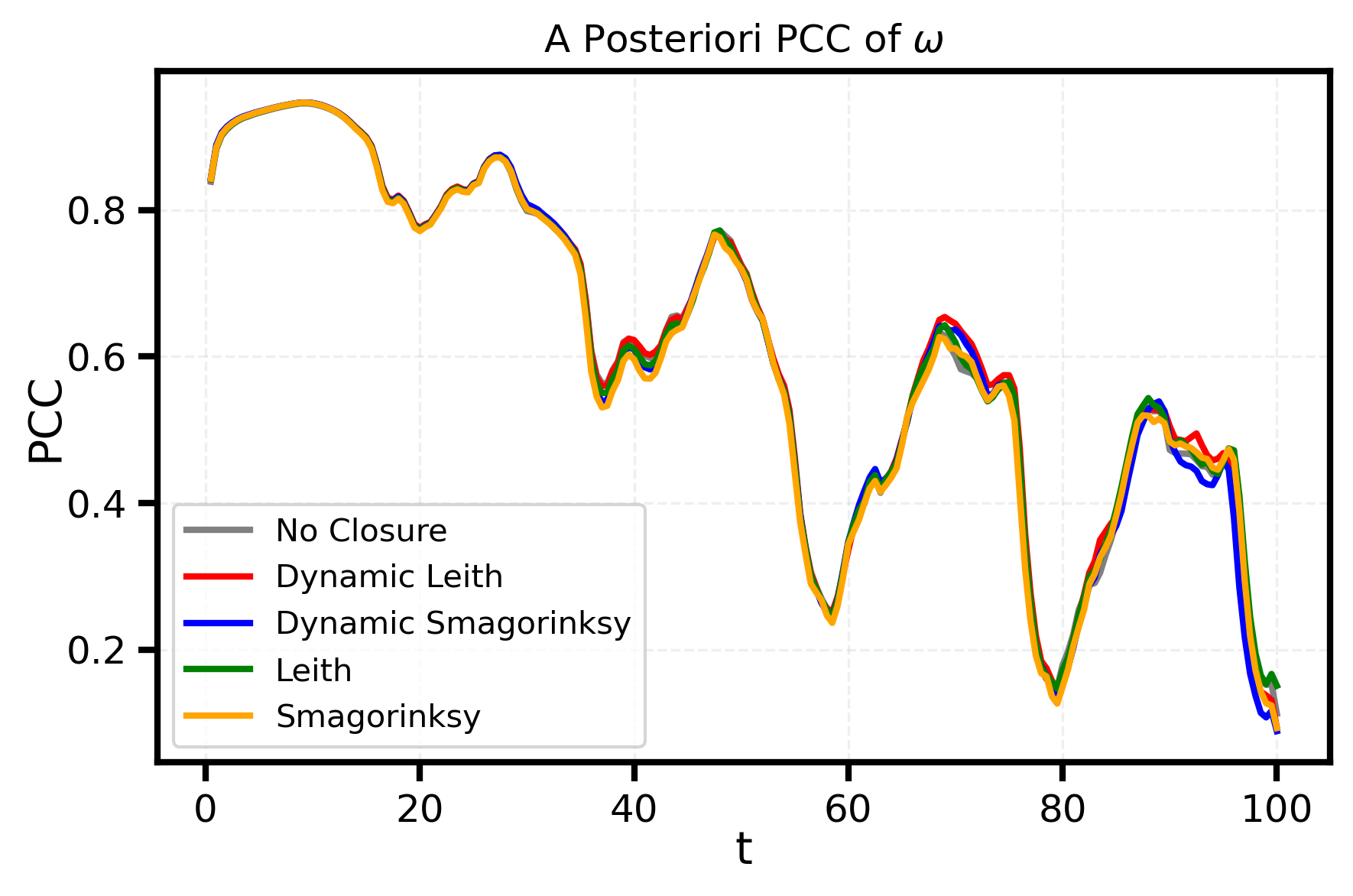}}
    \subfloat[][$\beta = 0.1$]
    {\includegraphics[width=0.35\linewidth]{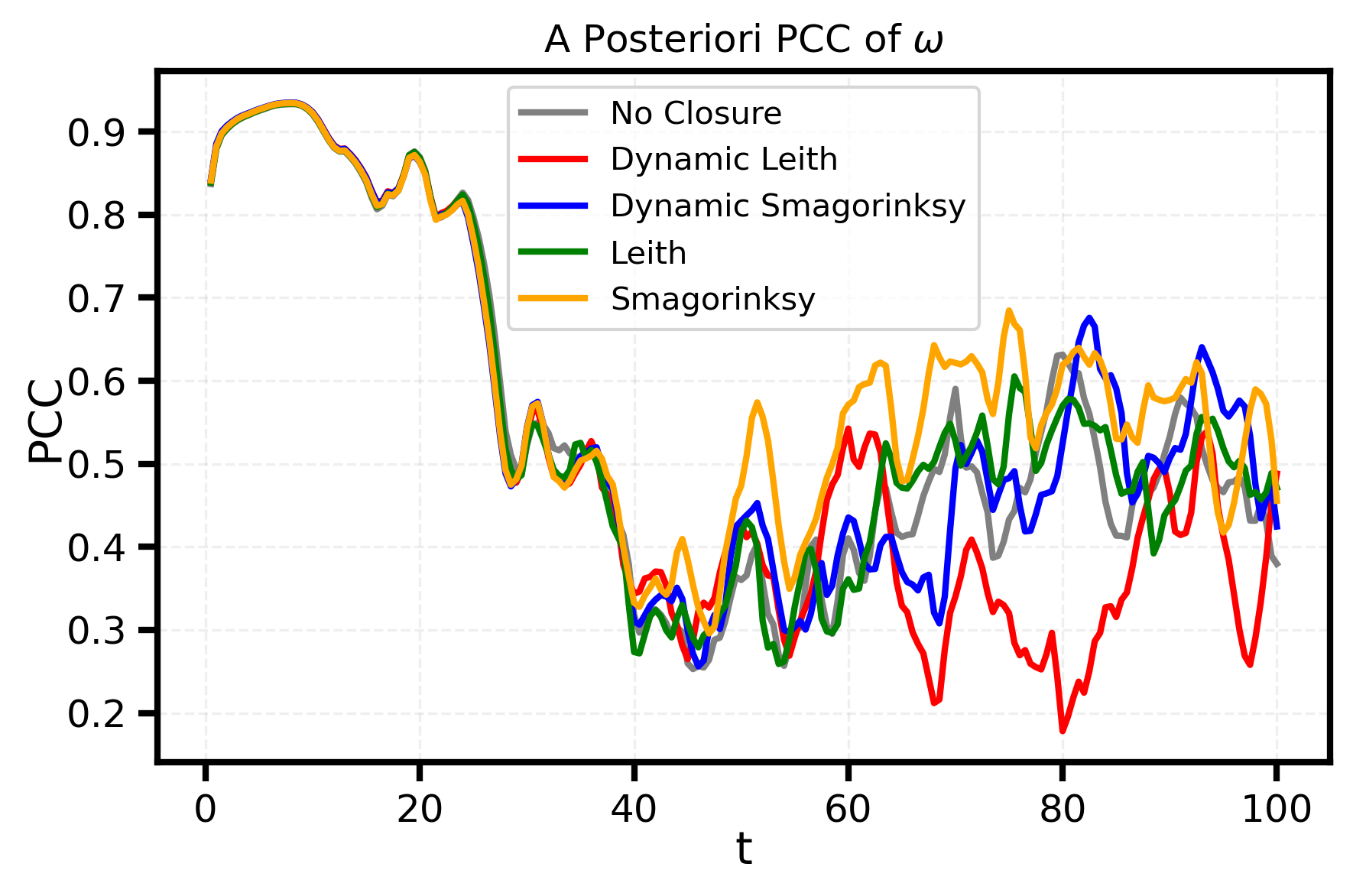}}
    \subfloat[][$\beta = 1$]
    {\includegraphics[width=0.35\linewidth]{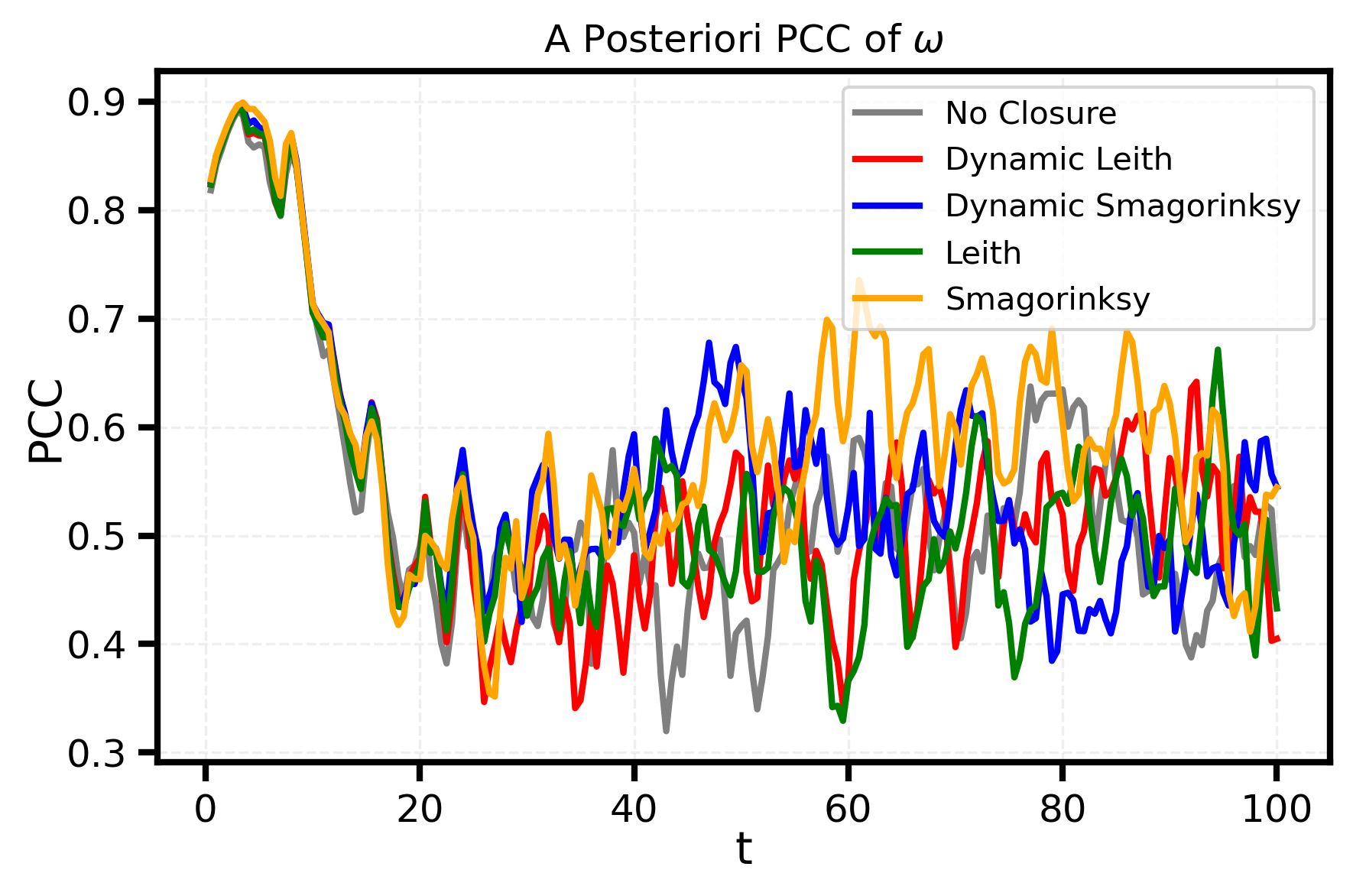}}
    \caption{ \small Evolution of a posteriori correlation $(\text{PCC}_{\text{a-post. }\omega})$ of $\omega$ for flows past an idealized cape with various $\beta$}
    \label{fig:aposteriori_cape}
\end{figure*}

\begin{figure*}[h]
    \centering
    \subfloat[][Filtered Field]
    {\includegraphics[width=0.35\linewidth]{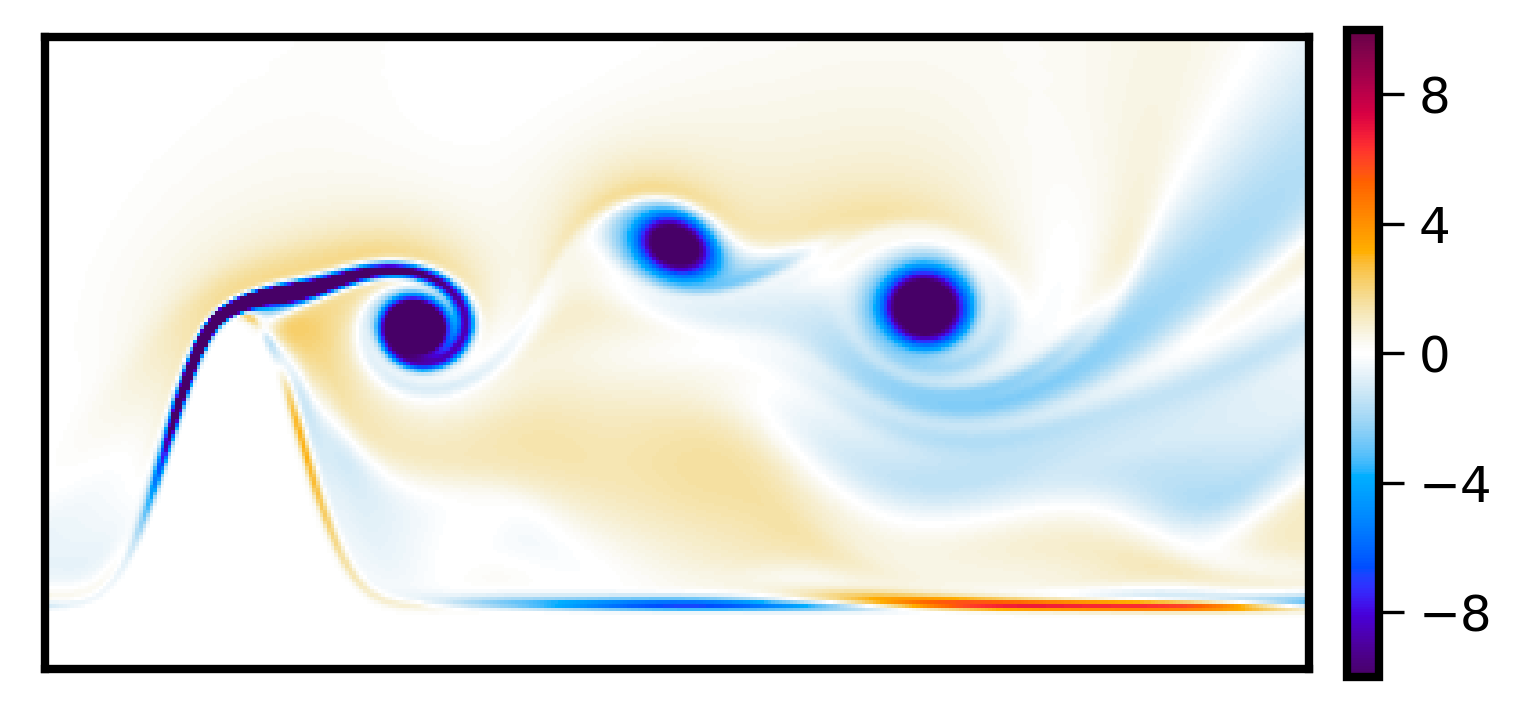}}
    \subfloat[][No Closure]
    {\includegraphics[width=0.35\linewidth]{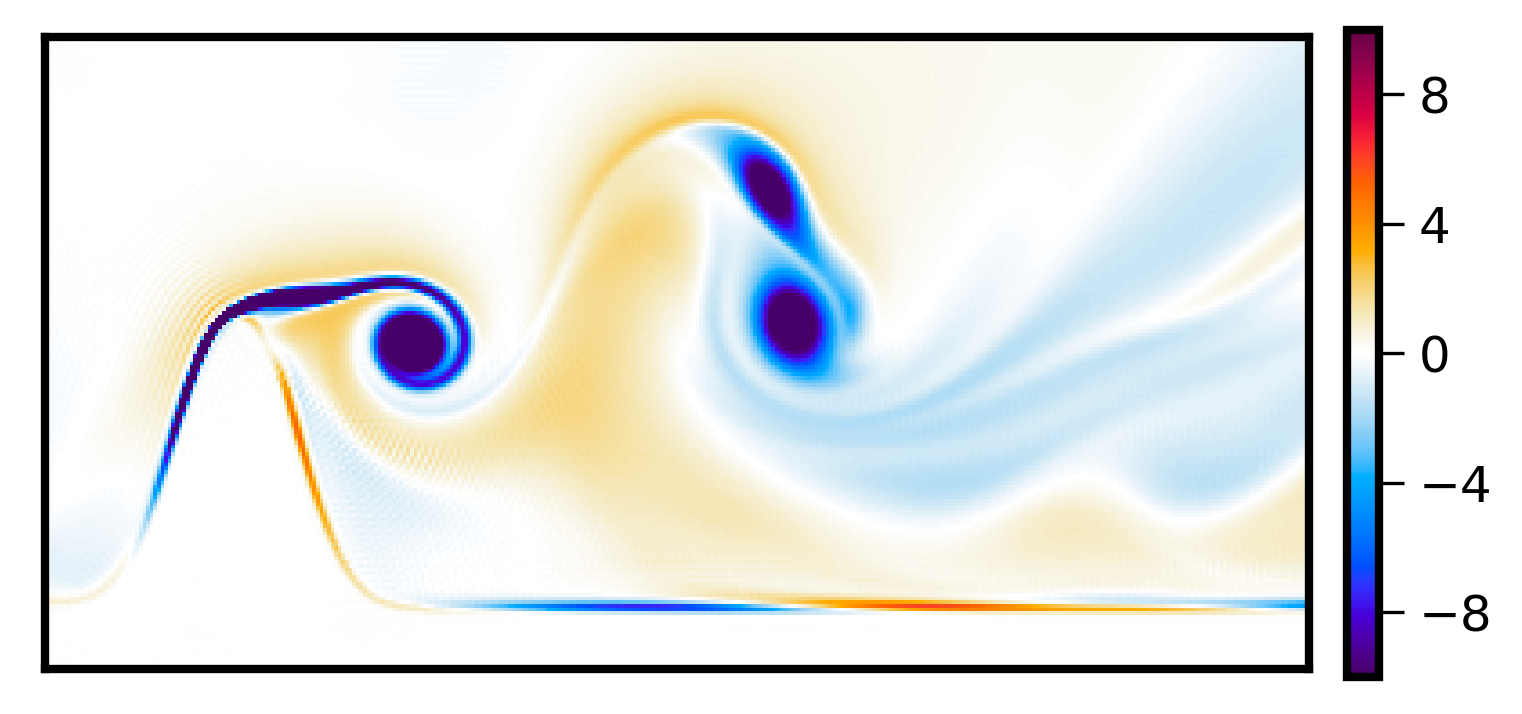}}
    \subfloat[][Dynamic Smagorinsky]
    {\includegraphics[width=0.35\linewidth]{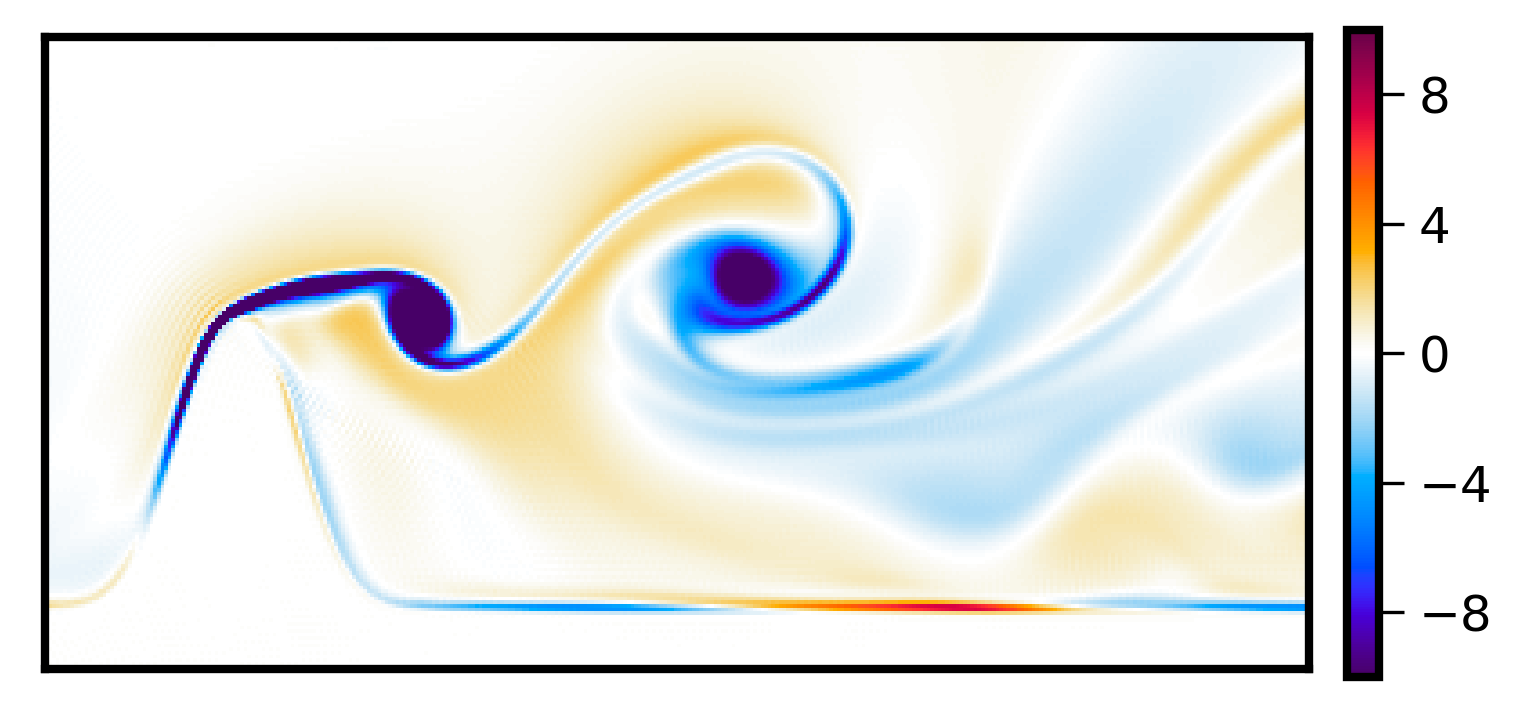}} \\
    \subfloat[][Dynamic Leith]
    {\includegraphics[width=0.35\linewidth]{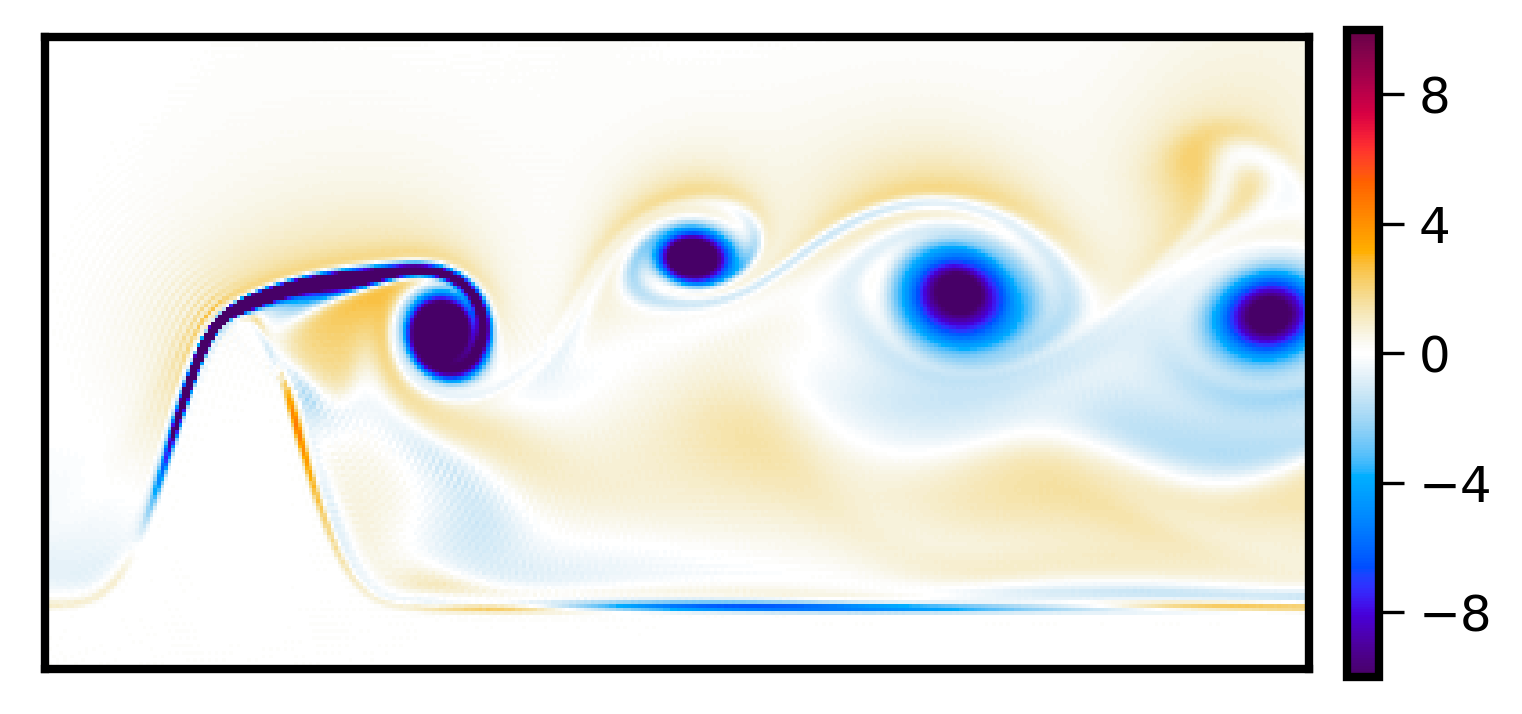}}
    \subfloat[][Smagorinsky]
    {\includegraphics[width=0.35\linewidth]{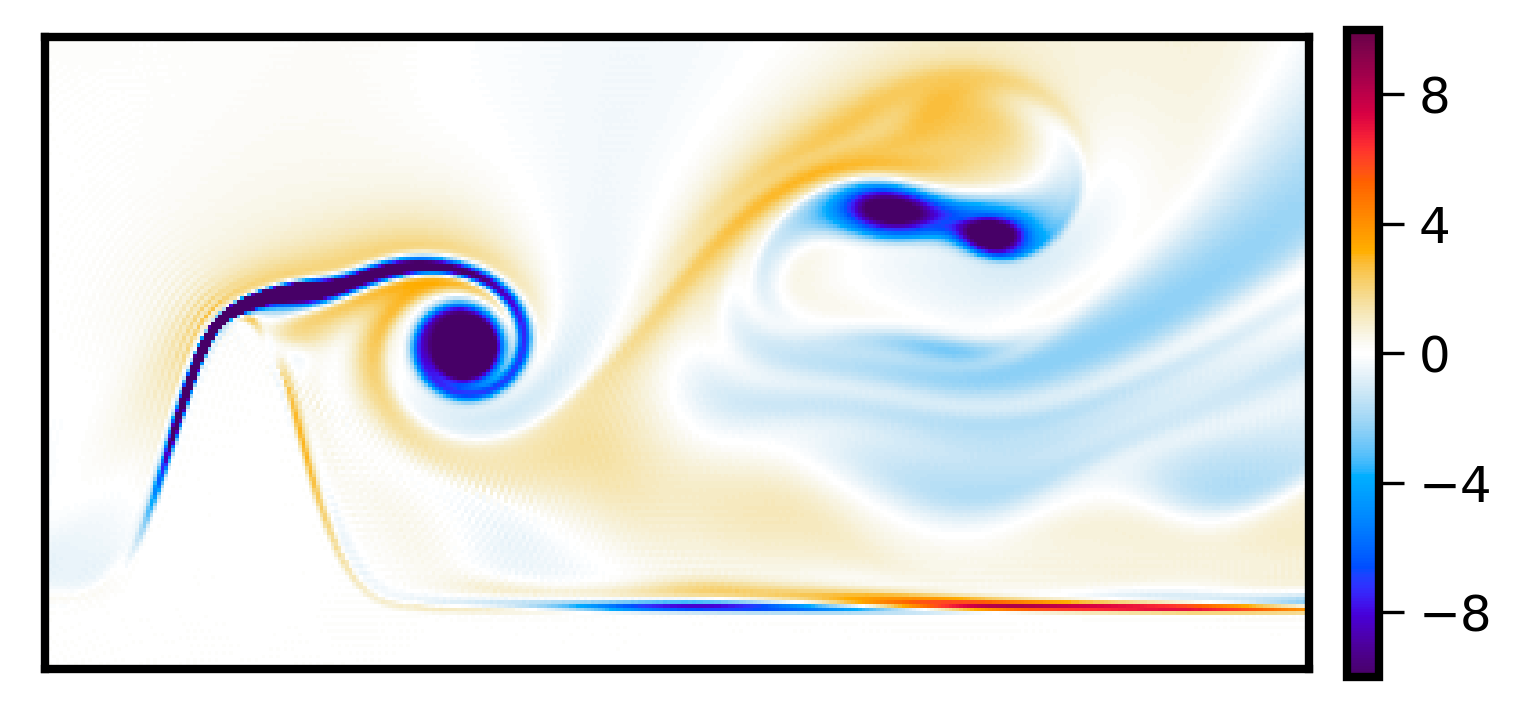}}
    \subfloat[][Leith]
    {\includegraphics[width=0.35\linewidth]{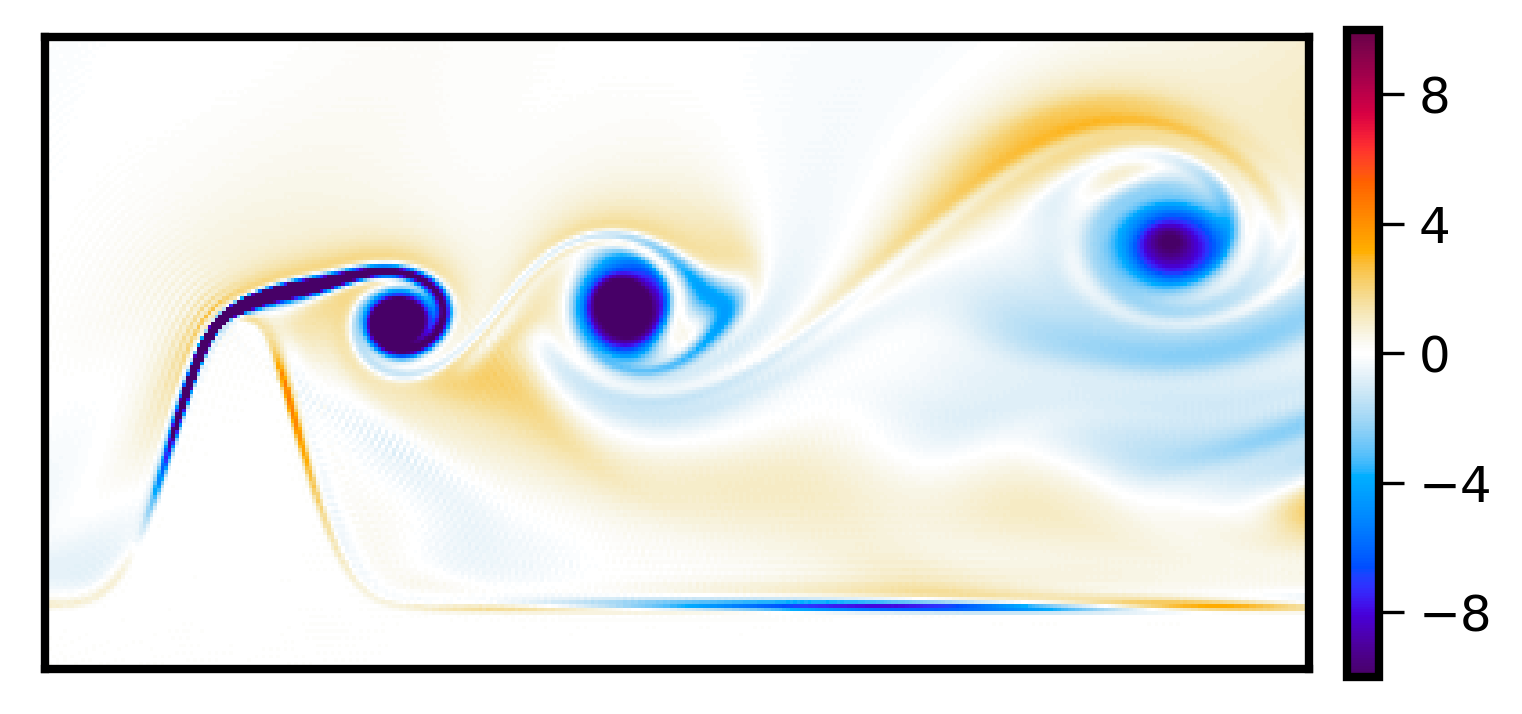}}
    \caption{ \small (a) Vorticity of the filtered field, and (b-f) vorticity of the LES most similar (minimum phase shift error) to the filtered field, for flows past an idealized cape with $\beta = 1$}
    \label{fig:field_cape}
\end{figure*}

 We vary $\beta \in \{0,0.1,1\}$ to focus on boundary effects and vortex shedding rather than statistical turbulence, as shown with fully-resolved fields in Fig.\;\ref{fig:DNS_cape}. The flow progresses from single-shed vortices to a von-Kármán vortex street and finally to a shedding zonal jet as $\beta$ increases.

As before, the LES closures are run with square grid size $512 \times 512$ ($\delta=2$), and timestep 10 times larger than that of the fully-resolved, $N_{FR} \times N_{FR} = 1024 \times 1024$ field.

\subsubsection{A Priori Analysis}

As before, in Table\,\ref{tab:Cape_priori_PCC} we compare a priori PCC ($\text{PCC}_\text{a-priori}$) for the analytical closures. Leith performs better than Smagorinsky, but due to richer flow features, the a priori performance is reduced compared to the flow past circular islands case. Both closures degrade slightly with higher Coriolis variation ($\beta$).

\begin{table}[]
    \centering
    \begin{tabular}{c|c|c}
          \hline
          Beta & (Dynamic) Smagorinsky & (Dynamic) Leith   \\
         \hline
         0 & 0.3284 & 0.3697  \\
        \hline
        0.1 & 0.3315 & 0.3694 \\
        \hline
        1 & 0.2980 & 0.3328 \\
        \hline
    \end{tabular}
    \caption{\small Time-averaged a priori correlation $(\text{PCC}_\text{a-priori})$ of $\Pi$ for flow past an idealized cape with various $\beta$}
    \label{tab:Cape_priori_PCC}
\end{table}

\subsubsection{A Posteriori Assessment}

In Fig.\;\ref{fig:SGS_cape}, we compare the SGS PCC ($\text{PCC}_\text{a-post. SGS}$) for the analytical closures with optimal $c$ as shown in Fig.\;\ref{fig:c_cape}. Now, dynamic Leith also performs better than Leith, while Smagorinsky and dynamic Smagorinsky again show similar PCC values. Temporal variation of the SGS PCC is extremely similar to the circular case: from Fig.\;\ref{fig:SGS_cape}, Smagorinsky shows higher initial correlation and then all closures oscillate with the vortex shedding. Due to the more complex cape geometry, the oscillations now have a unique signature (visible for $\beta=0$) that decorrelates as $\beta$ increases. The higher Coriolis variation induces background Rossby waves and jets, so SGS now differs at more than just the coastal boundary layer. Closure evaluation is thus biased toward the open ocean, where the Smagorinsky and Leith closures are well validated and thus perform better \cite{fox2008can}, opposite the simplistic a priori trend.
Both dynamic $c$'s also show peaks and troughs that follow vortex shedding patterns, and the oscillations increase in frequency with increasing Coriolis variation ($\beta$).
We now consider vorticity. The values of $c_S$ and $c_L$ from the SGS analysis no longer remain optimal for the reconstruction of the vorticity field. The optimal values obtained for the vorticity field are $c_L = 0.2, 0.15$ and $0.2$ with $\beta = 0, 0.1$ and $1$, respectively, and $c_S = 0.15$ for $\beta = 0.1$. 
\begin{table}[]
    \centering
    \begin{tabular}{c|c|c|c|c}
          \hline
          No Closure & Dyn. Smag. & Dyn. Leith & Smag. & Leith   \\
          \hline
        3 & 3.5 & 4.5 & 0 & 5 \\
        \hline
    \end{tabular}
    \caption{ \small Phase shift (in non-dimensional time units) between the filtered field and LES for flow past an idealized cape with $\beta = 1$}
    \label{tab:phase_cape}
\end{table}
Fig.\;\ref{fig:aposteriori_cape} shows that the simulated vorticity fields are correlated for longer times than in the cylinder test case for $\beta =0$ and $0.1$. As in the a priori and SGS analyses, there is a strong vortex-shedding signature, which again decorrelates and increases in frequency with increasing $\beta$. 
There is also an increase in overall vorticity PCC ($\text{PCC}_{\text{a-post. }\omega}$) with $\beta$, as before, due to the presence of zonal jets. 
Table\,\ref{tab:phase_cape} lists the phase-shift, which shows that all closures other than Smagorinsky are equivalent or worse in gross phase error than the unclosed LES. However, Fig.\;\ref{fig:field_cape} shows that in this application, there are also errors arising from incorrect or missing features such as eddies, and not just phase shift, which are captured well by the Leith and Dynamic Leith models.

\section{Conclusions and Discussion}

We extended existing pseudo-spectral quasi-geostrophic (QG) numerical schemes to simulate aperiodic flows with coastal boundaries. We used the Brinkman volume penalization approach to handle land masks and a splitting algorithm for sponging with inflows, outflows, and aperiodic boundary conditions. With this new GPU-based solver, we evaluated the performance of four subgrid-scale (SGS) closure models for Large Eddy Simulation (LES): the Smagorinsky model, the Leith model, and their dynamic variants. We showcased applications to periodic eddy shedding past a circular island and flow past idealized capes in the $\beta$-plane. A priori analysis indicated that correlations between the analytical closures and the true SGS closure are moderate for flow past circular islands, and the correlation becomes weaker with larger values of $\beta$ for flow past idealized capes. A posteriori assessments showed much weaker overall correlation, feature reconstruction and large phase-shift errors in the simulated LES fields, with the closures failing to capture the correct eddy shedding frequency in both applications. 

Future work could focus on evaluating structural closure models, such as the Gradient Model \cite{clark1979evaluation} or data-driven closures \cite{zhang2025addressing} with coastal boundaries. Since ideal LES is inherently stochastic \cite{langford1999optimal}, evaluation and development of stochastic models for closure \cite{lermusiaux_JCP2006, suresh_babu_et_al_JAMES2025,guillaumin2021stochastic} could also be useful.

\section{Acknowledgments}
We thank the members of our MIT-MSEAS group and ML-SCOPE MURI team for discussions. We also thank Andrew Horning (RPI/MIT) for discussions on pseudo-spectral methods. We are grateful to the Office of Naval Research for partial support under grant N00014-20-1-2023 (MURI ML-SCOPE) and N00014-24-1-2715 (DRI-RIOT) to the Massachusetts Institute of Technology. ANSB was partially supported by an MIT Mechanical Engineering MathWorks Fellowship.

\small{
\bibliographystyle{IEEEtranN}
\bibliography{mseas,references_paper}}

\end{document}